\documentclass[lettersize,journal]{IEEEtran}
\usepackage{amsmath,amsfonts,amssymb}
\usepackage[caption=false,font=normalsize,labelfont=sf,textfont=sf]{subfig}
\usepackage{textcomp}
\usepackage{stfloats}
\usepackage{url}
\usepackage{verbatim}
\usepackage{graphicx}
\usepackage{cite, doi}
\usepackage{orcidlink}
\usepackage{ulem}
\usepackage{soul}
\usepackage{overpic}
\usepackage{xparse}

\usepackage{todonotes}

\usepackage{booktabs, ulem, colortbl, makecell, mathtools, multirow}
\usepackage{bm}
\usepackage{tikz}
\usepackage{enumitem}

\AtBeginDocument{%
  \providecommand\BibTeX{{%
    \normalfont B\kern-0.5em{\scshape i\kern-0.25em b}\kern-0.8em\TeX}}}

\newcommand{\noise}[1]{
   \overset{\rotatebox{-90}{$\nmid$}}{#1}}

\definecolor{skyblue}{RGB}{86, 180, 233}
\definecolor{vermilion}{RGB}{213, 94, 0}

\title{Collaborative Problem Solving in Mixed Reality: \\ A Study on Visual Graph Analysis}

\author{
Dimitar Garkov\,\orcidlink{0000-0002-2134-1686},
Tommaso Piselli\,\orcidlink{0000-0002-7088-920X},
Emilio Di Giacomo\,\orcidlink{0000-0002-9794-1928},
Karsten Klein\,\orcidlink{0000-0002-8345-5806},\\
Giuseppe Liotta\,\orcidlink{0000-0002-2886-9694},~\IEEEmembership{Senior Member,~IEEE},
Fabrizio Montecchiani\,\orcidlink{0000-0002-0543-8912},
Falk Schreiber\,\orcidlink{0000-0002-9307-3254}
\thanks{Manuscript received xx x 202x; revised x xx 202x; accepted xx xxx 202x. Date of publication xx xxx 202x; date of current version xx xx 202x.
This work was supported by Deutsche Forschungsgemeinschaft (DFG, German Research Foundation) under Germany’s Excellence Strategy - \mbox{EXC 2117} - 422037984 and DFG project ID 251654672 - \mbox{TRR 161}; by MUR PRIN under Proj.~2022TS4Y3N - EXPAND: scalable algorithms for EXPloratory Analyses of heterogeneous and dynamic Networked Data and Proj.~2022ME9Z78 - NextGRAAL: Next-generation algorithms for constrained GRAph visuALization, and by Università degli Studi di Perugia, Fondi di Ricerca di Ateneo, edizione 2022 - MiRA: Mixed Reality and AI Methodologies for Immersive Robotics.
\textit{(Corresponding author: Dimitar Garkov)}}
\thanks{\textit{Dimitar Garkov, Karsten Klein, and Falk Schreiber are with the Department of Computer and Information Science, University of Konstanz, Konstanz, Germany. E-mail: [first].[last]@uni-konstanz.de}}
\thanks{\textit{Tommaso Piselli, Emilio Di Giacomo, Giuseppe Liotta, Fabrizio Montecchiani are with the Department of Engineering, University of Perugia, Perugia, Italy. E-mail: [first].[last]@unipg.it, and tommaso.piselli@dottorandi.unipg.it}}
\thanks{\textit{Falk Schreiber is with the Faculty of Information Technology, Monash University, Melbourne, Australia. E-mail: falk.schreiber@monash.edu.}}
\thanks{This article has supplementary downloadable material and data available at \url{https://doi.org/10.18419/darus-4231} provided by the authors.}
}

\IEEEpubid{\textit{\copyright~2026 IEEE}}

\usepackage{etoolbox}
\makeatletter
\pretocmd{\@maketitle}{%
\vspace*{-4ex}
\textit{\normalsize\copyright~2026 IEEE. This is the author's version of the article that has been published in IEEE Transactions on Visualization and Computer Graphics. The final, published article is available at \href{https://doi.org/10.1109/TVCG.2026.3671472}{\color{blue}10.1109/TVCG.2026.3671472}.\vspace*{-0.5ex}}}{}{}
\makeatother

\begin{document}

\maketitle

\begin{abstract}
Problem solving is a composite cognitive process, invoking a number of cognitive mechanisms, such as perception and memory. 
Individuals may form collectives to solve a given problem together in collaboration, especially when complexity is perceived to be high. 
To determine if and when collaborative problem solving is desired in the context of visual graph analysis, we compare ad hoc pairs to individuals and nominal pairs, when solving different tasks in mixed reality. 
We discuss the results of an experiment with 72 participants performed in two countries and three languages. 
We apply the concept of task instance complexity to quantify the visual demand of tasks used in the experiment. 
Our results show the importance of using nominal groups as a benchmark for evaluating collaborative virtual environments. 
We conclude that 3D graph representation is not sufficient to induce better collaborative results compared~to the benchmark. 
\end{abstract}



\begin{figure*}[t]
    \centering
    \hspace*{-0.22cm}\subfloat{
    \begin{overpic}[abs,unit=1mm, width=.25178069353\linewidth, height=5.8cm]{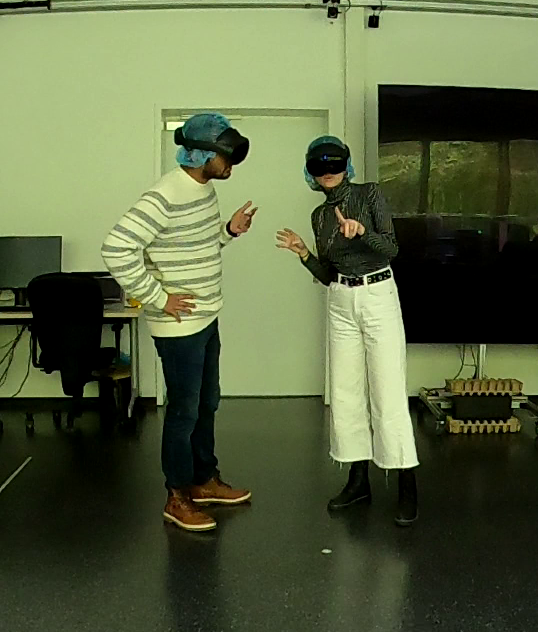}%
    \put(21.875,3){\textcolor{white}{\small\bfseries (a)}}%
    \end{overpic}}%
    \hspace*{-0.15cm}\subfloat{
    \begin{overpic}[abs,unit=1mm, width=.38699625117\linewidth, height=5.8cm]{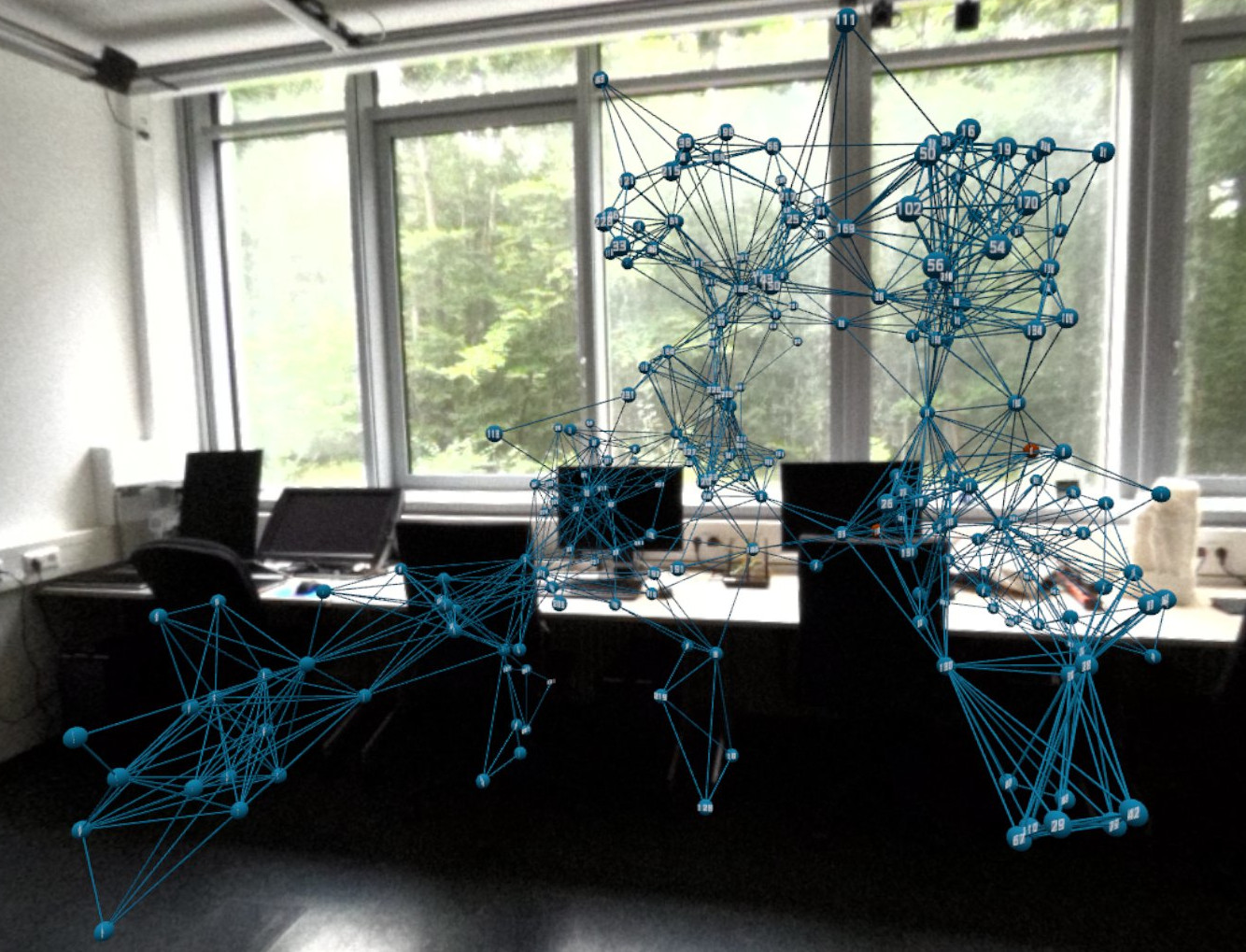}%
    \put(33.25,3){\textcolor{white}{\small\bfseries (b)}}%
    \end{overpic}}%
    \hspace*{-0.19cm}\subfloat{
    \begin{overpic}[abs,unit=1mm, width=.17811152764\linewidth, height=5.8cm]{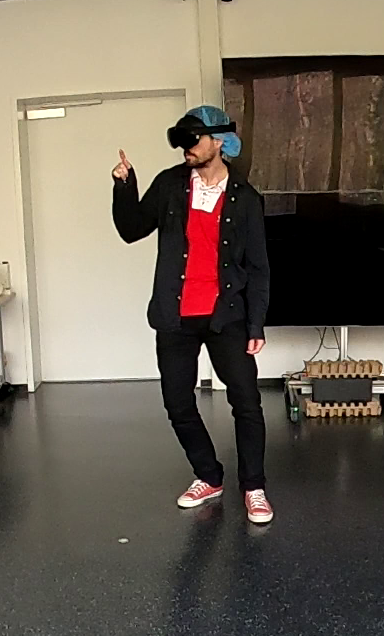}%
    \put(13.875,3){\textcolor{white}{\small\bfseries (c$_1$)}}%
    \end{overpic}}%
    \hspace*{-0.18cm}\subfloat{
    \begin{overpic}[abs,unit=1mm, width=.17811152764\linewidth, height=5.8cm]{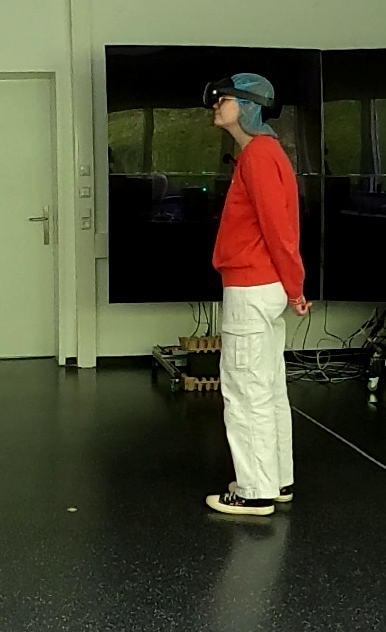}%
    \put(13.875,3){\textcolor{white}{\small\bfseries (c$_2$)}}%
    \end{overpic}}%
    \caption{Participants solving visuospatial tasks on graphs in mixed reality~(b).
    Participants who were assigned to ad hoc pairs~(a) solved the tasks collaboratively, while the remaining participants solved these on their own, either as individuals~(c$_1$ or c$_2$) or in a nominal pair~(c$_1$ and c$_2$).}%
    \label{fig:main}%
\end{figure*}

\section{Introduction}
\IEEEPARstart{W}{e} are surrounded by situations that involve problem solving.
Sensation, perception, memory, and inference are invoked to let us identify, search, memorize, or abstract~\cite{wang2006layered}.
We apply strategies, such as using heuristics and analogies, to tasks of different complexity~\cite{wood1986task, campbell1988task} and in collaborative contexts.
Collaboration itself is characterized by the interaction among group members, e.\,g., by bringing members on the same page~\cite{clark1991grounding} or in devising effective strategies~\cite{schwartz1995emergence}.
During collaboration, however, group interaction also incurs process costs~\cite{steiner1972group, clark1991grounding}.
If process costs can be overcome, the benefits can be expected to increase with task complexity~\cite{shaw1976assembly,wolf2013accurate}.

Like other visuospatial mediums~\cite{milgram1995augmented}, mixed reality~(MR) is predominantly associated with stereoscopic head-mounted displays~(HMD), with the first such display built in 1968~\cite{sutherland1968head}. 
Since then, display technology and our understanding of it have advanced significantly, extending to fields, such as immersive analytics~\cite{marriott2018immersive}, where collaboration has been identified as a \textit{grand challenge}~\cite{ens2021grand}.
Previous studies involving collaborative graph analysis have targeted factors, such as the environment~\cite{cordeil2016immersive}, interaction~\cite{prouzeau2016evaluating}, and symmetry~\cite{tong2023towards}.
A survey~\cite{huang2022learning} covering collaborative problem solving, learning spatial skills, and virtual reality~(VR) categorized studies as having three foci: applications of VR, virtual factors supporting collaboration, and physical factors supporting collaboration; other surveys~\cite{ens2019revisiting, sereno2020collaborative} draw attention to factors, such as presence~\cite{steuer1992defining} and awareness~\cite{dourish1992awareness}.
This highlights a strong trend of focusing research \textit{in}~rather~than~\textit{on} collaboration that stands in contrast with traditional \mbox{research} on groups.
Groups merit more systematic studies that support  the growing body of literature on collaborative virtual environments (CVE).
Only after benchmarking collaboration can fundamental results reliably inform applications in CVE.

This study focuses on collaborative problem solving in a visuospatial, collocated MR setting~(Fig.~\ref{fig:main}).
Most studies on CVE do not probe collaboration itself, and the few that do~\cite{heldal2006two, garcia2007collaborative, sando2011impact, prouzeau2016evaluating} take place in CVE different from stereoscopic \mbox{3D} and omit to control for group size effects.
Apart from group size, the type of collaborative group is also controlled.
This is the first such study in CVE devoted to visual graph analysis.
It is worth recalling that visual graph analysis is particularly applicable to problem solving in 3D, where locating, orienting, navigating, and understanding of spatial objects and their relationships are invoked concurrently~\cite{huang2022learning}.
\IEEEpubidadjcol

\textbf{Contributions.}
While being in pairs~(Fig.~\ref{fig:main}~(a)) or on their own~(Fig.~\ref{fig:main}~(c$_1$),~(c$_2$)), participants were asked to complete two commonly used tasks in visual graph analysis:~1)~count the number of common neighbors; and 2)~determine the length of the shortest path.
To examine the effects of each task trial, we define \textit{task~instance~complexity}.
Specifically, we address the following~\mbox{research~questions}:

\begin{itemize}
    \item[\textbf{{\scriptsize RQ}{\small 1}}] Do accuracy and completion time of any pair differ from that of one individual or two individuals?
    \item[\textbf{{\scriptsize RQ}{\small 2}}] Relatedly, do differences in accuracy and completion time change with task instance complexity and how?
    \item[\textbf{{\scriptsize RQ}{\small 3}}] Does teamwork affect cognitive workload differently when total task instance complexity is high or low?
\end{itemize}

Our contributions can be summarized as follows:

\begin{itemize}
    \item[--] a controlled experiment in collaborative problem solving in MR applying visual graph analysis, that was conducted in two countries and three languages,
    \item[--] an investigation into how (collaborative)~problem solving scales for different levels of task instance complexity,
    \item[--] an analysis of the differences in solving strategies and the sense of flow in pairs and individuals,
    \item[--] a benchmark for disentangling collaboration from \textit{two heads are better than one} effects in pairs for CVE.
\end{itemize}

Our study suggests that collaborating around a given \mbox{3D} graph rep\-resentation is not sufficient to achieve results better than those of an equal number of persons working~independently.
Increasing task instance complexity is also generally not sufficient to change this state of affairs.
Results from the study may serve~to benchmark collaborative graph analytics without advocating against it.
To support replication of the study, we make data and source code openly available~\cite{darus-4231_2024}.

\section{Related Work}
In the following, we focus on results from controlled experi\-ments, but include other results to capture the state-of-the-art.

\subsection{Group problem solving}
In behavior science and social psychology, groups have been studied for most of the past hundred years, where a group is defined as~\cite{hackman2012causes}:

\begin{quote}
\textit{A group is an intact social system, complete with boundaries, interdependence for some shared purpose, and differentiated member roles.}  
\end{quote}

Group problem solving, as measured by time and score, depends on the task at hand~\cite{thorndike1938effect, hackman1969toward}.
For example, groups require more time per person~\cite{husband1940cooperative, taylor1952twenty}, but tend to achieve higher levels of abstraction~\cite{schwartz1995emergence} and have higher completion rates than individuals~\cite{thorndike1938effect, marquart1955group, faust1959group}.
The strength of groups lies more in the group's self-correcting mechanism~\cite{thorndike1938effect} rather than the number of contributions~\cite{marquart1955group}.
Such findings stress the importance of evaluating groups against equally sized pools of individuals~\cite{marquart1955group, faust1959group, hill1982group}.
Group effectiveness is inferior to pooled effectiveness, but it may be hypothetically veered through the interaction among group members~\cite{hackman1975group}
given that the course of action in the group is often \mbox{suboptimal~\cite{steiner1972group, hill1982group, kerr2004group}}.
Researchers have therefore searched for so-called \textit{assembly bonus \mbox{effects}}~\cite{shaw1976assembly, kerr2004group} such that groups can outperform both individuals and pools of individuals.
Task complexity may be one factor~\cite{shaw1976assembly}, but evidence has been scarce.
More recently, Hackman and Katz~\cite{hackman2010group} formulated questions for evaluating group effectiveness
and identified the group attributes responsibility, synchronicity, authority, and work type. 
As research opportunities, they outlined investigating established norms for new technologies and evaluating their usefulness.
To this end, Wikst\"om et al.~\cite{wikstrom2020collaborative} developed and evaluated a task for collaborative problem solving in VR and reality, that was a 3D puzzle with shapes.

\subsection{Collaboration in virtual environments}
Collaboration in virtual environments can be collocated or distributed
and synchronous or asynchronous~\cite{billinghurst2018collaborative}.
Compared to VR, collocated collaboration in MR benefits from face-to-face communication~\cite{billinghurst2002collaborative}. 
It can preserve non-verbal cues~\cite[Exp.\,1]{kiyokawa2002communication}
and serve well in communicating common ground~\cite{benk2022my} or spotting errors~\cite{wang2011comparative}.
In collocated collaboration, pairs also tend to prefer a same-side perspective~\cite{poretski2021physicality} for which dedicated techniques, such as supporting shared awareness~\cite{bovo2022cone}, have started to emerge.
As for collaborative navigation, early results favored egocentric over exocentric perspectives~\cite{yang2002exploring}.
Spatial referencing, which can help reduce process loss~\cite{steiner1972group}, was discussed by Heer and Agrawala~\cite{heer2007design} who identified nuanced pointing as one design consideration for improving collaboration.
In terms of space use, participants (individually or in small groups) seem to prefer utilizing different objects in the physical environment, e.\,g., walls~\cite{lee2020shared}, whiteboards or tables~\cite{luo2022should}. 
Without physical affordances, the use of space is more mixed, especially when working alone~\cite{batch2019there}.

Few studies have scrutinized the value of collaboration in virtual environments.
Comparing pairs to individuals, differences were found for earlier types of environments~\cite{heldal2006two, garcia2007collaborative}, interaction techniques~\cite{prouzeau2016evaluating}, and small group sizes~\cite{sando2011impact}.
Using propagating selection, pairs made less errors in finding the shortest path and were faster than individuals when complexity was high~\cite[Exp.\,1]{prouzeau2016evaluating}.
In Sando et al.'s experiments, pairs were more accurate than individuals and groups of four in judging relative distances~\cite[Exp.\,1]{sando2011impact}, unless several perspectives were required~\cite[Exp.\,2]{sando2011impact}.
Individuals were faster overall, unless perspectives were part of the task, then the groups of four were faster~\cite[Exp.\,3]{sando2011impact}.
There are no studies that have separated collaboration from aggregation in virtual environments, e.\,g., comparing against nominal groups~(see Sec.~\ref{defs}).

\subsection{Collaborative visual graph analysis in virtual environments}
During collaboration, group performance is affected by the visualization and its interactivity~\cite{bresciani2009benefits} or stereo\-scopicity~\cite{narayan2005quantifying}.
Many taxo\-nomies exist for graphs, e.\,g., for common low-level tasks~\cite{lee2006task}, mental-map tasks~\cite{archambault2013map}, and temporal graphs~\cite{kerracher2015task}.

``Collaborative graph analysis is not represented prominently''~\cite{von2011visual}
and there is not much empirical evidence from CVE~\cite{prouzeau2016evaluating, cordeil2016immersive, tong2023towards}.
Prouzeau et al.~\cite[Exp.\,1]{prouzeau2016evaluating} validated the aforementioned selection technique on large interactive displays with graphs of different size and density that were planar and small-world.
Cordeil et al.~\cite{cordeil2016immersive} proved the utility of HMD-mediated CVE over less affordable CAVE environments.
The authors used two global tasks:~1) find the shortest path; and 2) count the number of triangles; in random graphs of varying size but equal density.
Tong et al.~\cite{tong2023towards} evaluated asymmetric CVE and found their effectiveness to be on par with VR and non-VR environments.
Alongside a document-reading task, pairs drew node-link diagrams as solving aids.

A parallel but separate line of research is found in the field of collaborative learning---learning with graph-oriented representations.
There, knowledge maps were found consistent with higher hypothesis generation, consensus reaching, and integration of information, compared to threaded discussions~\cite{suthers2008beyond}.
Studies have also explored the benefits of representations, such as concept maps~\cite{gijlers2013using} and argumentative diagrams~\cite{van2007representational}.

\section{Definitions}\label{defs}
Here we introduce the formal terms we refer to in this paper.

\paragraph{\textbf{Individual}} a person who solves tasks alone. 

\paragraph{\textbf{Ad hoc pair}} two persons who solve tasks together. 
The term \textit{ad hoc} implies the pair is formed for a specific pur\-pose and it is not practiced.
Prac\-tice may affect collaborative task solving\cite{fox1962relative, taylor1952twenty} for which ad hoc pairs are preferred.
Working in an ad hoc pair is what we refer to as \textit{collaboration}.

\paragraph{\textbf{Nominal pair}} two persons who solve tasks independently, but whose results are evaluated together.
Nominal groups~\cite{steiner1972group, hackman1975group, hill1982group, hackman2010group} have their uses, e.\,g., for generating more new ideas~\cite{hill1982group} or in reference to wisdom of crowds~\cite{galton1907vox, kameda2022information}.
We employ this technique to differentiate collaborative from aggregate contributions.

\paragraph{\textbf{Task complexity}}
We define a task as \textit{complex}, if the task's composition of \textit{task attributes}~\cite{wood1986task, hackman1969toward} consists of heterogeneous subcompositions~\cite{kim2008complex, yoghourdjian2020scalability}~(not only a series of subtasks~\cite{lee2006task}) and the \textit{behavior requirements}~\cite{wood1986task} to complete the task are high.
We use this concept to define \textit{task instance complexity}.

\paragraph{\textbf{Task instance}} A task can be represented differently. 
For a given task, e.\,g., finding the shortest path in a graph, and its representation, e.\,g., a node-link diagram, there are one or more distinct \textit{task instances}.
A task instance is a specific visual configuration on which the task is executed.
Different task instances may affect task solving differently.
    
\paragraph{\textbf{Task instance complexity}}
is measured by looking at these two factors:
\begin{enumerate}[leftmargin=8ex]
    \item[1)] the interactions with the task instance using a reference solving strategy,
    \item[2)] the noise in the region of inspection~(ROI).
\end{enumerate}

With 1) we estimate the necessary interactions with task attributes on a path to the correct solution~\cite{campbell1988task}. 
The ROI may also contain noise, with the intuition that increased noise leads to costlier processing of attributes~\cite{wood1986task}, e.\,g., regarding search, identification, and memorization~\cite{wang2006layered}, or lack of certainty~\cite{sacha2015role}.
Task instance complexity is thus a task-specific measure for the lower bound of necessary interactions under some noise constraint.
We call 1) and 2):~\textit{signal} and \textit{noise} instance~complexity, respectively.

Other studies have pursued the opposite approach:~from~a set of known metrics, test the predictive power based on ob\-served data.
Yoghourdjian et al.~\cite{yoghourdjian2020scalability} apply dimensionality reduction to construct what they deem ``task hardness''.
Two classes of contributors seem evident~\cite[Fig.~9]{yoghourdjian2020scalability} that, along with dis\-par\-i\-ties across graph sizes and densities~\cite{ware2002cognitive, huang2009graph, yoghourdjian2020scalability}, would support the dual composition of task instance complexity.

\subsection*{Graph preliminaries}
An \emph{undirected graph} $G$, or network, is a tuple $G = (V, E)$ of a set of vertices, or nodes, $V$, and a set of edges, or links, $E \subseteq V \times V$. 
A \emph{drawing} $\Gamma(G)$ of a graph $G$, also called a \emph{layout} of $G$, maps every vertex of $G$ to a distinct point in the Euclidean space $\mathbb{E}^3$.
Additionally, every node and every edge has visual attributes, such as size, color, and edge type attached to it. 
Let $deg(u)$ be the \emph{degree} of $u$, i.\,e., the number of adjacent nodes, or neighbors, to $u$. Let $dist(u, v) = ||v - u||$ be the Euclidean distance from $u$ to $v$ in $\Gamma(G)$. For consistency, we use geodesic path~\cite[Fig.~2]{huang2009graph} to refer to $dist(u,v)$, and shortest path to refer to the shortest topological distance in $G$~(see Sec.~\ref{task2}).

\section{Experiment}
Next, we describe the experiment's design, tasks, task instance complexity methodology, hypotheses, and procedure.

\subsection{Experimental design}\label{exp-design}
The experiment follows a mixed design with two independent variables:~\textit{group type}~(ad hoc pairs, individuals, and nominal pairs), and \textit{task instance complexity}, and three dependent variables:~accuracy~(percentage), completion time~(seconds), and weighted workload ratings from NASA-TLX~\cite{hart1986nasa}~\mbox{(0--100).}
We remark that NASA-TLX is predominantly used  to measure subjective cognitive workload in HCI~\cite{kosch2023survey}.
NASA-TLX, however, has a broader scope and its dimensions are often analyzed separately; we report these in our Suppl.~Mat.~\cite[Fig.~S35]{darus-4231_2024}.

\subsubsection{Between-subject design}~\label{between-subject rationale}
For the independent variable group type, the design is between-subject, with nominal pairs invoked to differentiate collaborative from aggregate effects.
To create the pairs, we randomly assign half of all participants in the nominal pairs and half---in the ad hoc pairs.
Then, from each nominal pair we randomly nominate one \textit{individual}.
Comparisons are performed entirely against the ad hoc pairs.
Selecting individuals from the nominal pairs, as opposed to constructing nominal pairs from the individuals, avoids oversampling the individuals and ensures independence between the nominal pairs.

To harmonize the nominal pairs, we use the following setup.
Imagine we place each member in a separate room. 
We record as for the ad hoc pairs, i.\,e., assuming a parallel start.
In this scenario, we have to wait for all members to complete, before we can compare their results and pick the best one.
This is alternative to picking the first-to-complete member, regardless of the result.

To harmonize the responses of the ad hoc pairs, members are required to reach a consensus. 
Group consensus is important for decision making~\cite{dyer2009leadership} and naturally emerging in small groups~\cite{sherif1935study}. 
In case of no consensus, consensus is technically enforced as part of the interface---members cannot proceed until consensus is reached.

\subsubsection{Within-subject design}
For task instance complexity as our second independent variable,
the design is within-subject and repeated-measures.
We selected two distinct graph tasks as to not inhibit generalizability compared to a single task~\cite{purchase2012experimental}.
Introducing more tasks is, however, difficult due to the onset of discomfort during prolonged HMD use.
This is especially true for novice users and in line with prior studies~\cite{cordeil2016immersive}. 
The number of instances per task is selected such that a procedure lasts around an hour~\cite{rebenitsch2021estimating}, with no breaks for the sake of con\-trol\-la\-bil\-i\-ty.
Each participant gives $25$ answers:~$12$ instances of each task plus one control instance.
We counterbalance the order of the tasks and randomize the task instances.
The randomization of task instances is performed using the answer value and after controlling for graph properties.

\subsection{Tasks}\label{tasks}
To select the two tasks, we consulted several graph task taxonomies~\cite{lee2006task, burch2020state} and considered common tasks used in prior user studies on collaboration~\cite{cordeil2016immersive, prouzeau2016evaluating} and cognitive complexity~\cite{yoghourdjian2020scalability}.
We sought divisible but not pre-divided tasks to strengthen ecological validity.
With no predefined roles, the ad hoc pairs have to decide how to allocate and parallelize their resources.
Coordination can be expected to increase, but this is to avoid benefiting the ad hoc pairs.
Our task selection also distinguishes between a topologically local and a topologically global task.
Specifically, we asked participants to:
\begin{enumerate}
    \item[(1)] count the number of common neighbors for two selected nodes,
    \item[(2)] determine the length of the shortest path between two selected nodes,
\end{enumerate}
where the selected nodes are visually highlighted.

In posing the tasks, we control the environment, the graph, the layout, and certain graph properties.
For each task we remove outliers that may severely imbalance work division.
We discuss the task instance complexity methodology of each task separately in Sec.~\ref{task1} and Sec.~\ref{task2}.

\subsubsection{Environment}
Pair members involve each other more when face to face than side by side~\cite{kiyokawa2002communication}, which is beneficial while receiving instructions~\cite{pouliquen2015virtual}. 
The ad hoc pair's starting position is set as facing across the task space after which participants can move freely.
Each graph is placed in the center of an empty $9m^2$ area without any physical arrangements as to avoid potential effects on participants' interactions and movement.
The bounding box of each graph is uniform~($1m^3$) and equidistant to each starting point.
Its center is set to $1.45m$ from the ground. 
To not hinder verbal communication,
nodes are labeled with numbers.
The labels are placed centrally on the nodes and orientated cylindrically outwards from the view's center.

Graphs are drawn in the same slightly emissive sky blue color~\colorbox{skyblue}{\phantom{$\bullet$}}, whereas selected nodes---in vermilion~\colorbox{vermilion}{\phantom{$\bullet$}}.
The colors belong to the Okabe-Ito CUD schema~\cite{okabe2008color}
and were selected to accommodate color blindness, as well as to achieve salient contrast to the surrounding environment.
The lighting in the rendered scene follows a simple gradient model with emitters from above, below and the sides.
Bright sunlight is dimmed to not interfere with the headset's tracking or rendering. 

Participants can type their answers into a small keyboard interface using optical hand tracking.
The ad hoc pair's inputs are additionally synchronized.
To minimize interruptions in task solving due to movement, the interface stays with each participant~(in front, slightly tilted, at about chest height), unless this would interfere with the graph in which case it does not cross into the graph view.
A progress bar is shown once an instance is completed.
No other interactions are provided to avoid confounding due to tracking imprecision and interaction.

\subsubsection{Graphs}\label{netvars}
Social networks are known to be globally sparse and locally dense~\cite{henry2007nodetrix}, while also of practical relevance.
We selected real-world social networks from epidemiology, depicting the interactions of small mammalian species, known as field voles~\cite{davis2015spatial, sah2019multi}.
The dataset contains similar networks with topological properties not readily found in a probability model, such as a random, modular, or small-world model.
Graph mean size was $149$ nodes~($SD = 40$) and mean density was $0.05$~($SD = 0.01$).
Based on empirical evidence~\cite{yoghourdjian2020scalability}, we opted for larger, denser, and more varied graphs to have task instances of different complexity, to limit ceiling effects in group types, and to study interaction effects~\cite{shaw1976assembly, kerr2004group}.
Task instances were randomly drawn from $34$ graphs, where each graph is seen at most once by any participant.

\paragraph{Layout}
Graph layouts in 3D can be computed by transferring 2D methods to \mbox{3D} or by projecting and extending to \mbox{3D}.
We consider single-level energy-based methods where the layout results from the topology:~a force-directed layout~\cite{fruchterman1991graph} and stress minimization~\cite{kamada1989algorithm}.
For the force-directed method~\cite{fruchterman1991graph}, the extension to \mbox{3D} resulted in \textit{boxing} along the view's bounds~(see Suppl. Mat.).
A custom variant without a strict cut-off during simulation and with uniform normalization of results still had dense subgraphs drawn closely and even some node-node overlaps. 
We think an artifact-free distribution of the nodes requires a different choice of $k$, the ideal edge length~\cite{fruchterman1991graph}, which would near the node distribution of stress minimization.
Hence, we used stress minimization in 3D from OGDF~\cite{chimani2013open}, post-removal of smaller graph components.

\paragraph{Graph properties}
We cover global and local graph properties as follows. 
First, we compute all node candidates per task and their 1-neighborhood subgraphs~(not necessarily induced).
Then, we compute the graph properties to control for topology and geometric representation. 
Topologically, we compute local density, local clustering coefficient, degree centrality~(Task~1), and path betweenness~(Task~2); geometrically, we compute Euclidean node distance and fill ratio~(Suppl. Mat., Eq.~S1).
For each graph, we remove outliers whose z-scores lie at least two standard deviations from the mean.

\subsubsection{Task Instance Complexity of Task~1}\label{task1}
Based on the~task's topology~(\textit{Common~neighbors}), a minimum number of inspections are required to complete the task. 
One has to look at each neighbor and then inspect this neighbor's incident edges.
Let $u, v$ be the two selected nodes, we denote by $N_{uv}$ the set of common neighbors of $u$ and $v$.
Let $w$ be a common neighbor of $u, v$ such that $u \neq w \neq v$.
Let $E_{uv}$ be the set of edges connecting $u$ and $v$ to their common neighbors.
To reduce the number of inspections, one starts from $u$, assuming $deg(u) \leq deg(v)$.
This strategy focuses on the lower-degree node.
A competing strategy is to rely on the layout $\Gamma(G)$ and discretize the search space.
Assuming a uniform ROI, the likelihood that a ROI contains $w$ is proportional to its closeness to $u,v$.
To define ``closeness'', we could estimate the geodesic distance to the closest point $o$ to both $u, v$, which is the mid-point of the segment $\overline{uv}$.
In fact, traversing in both strategies is estimated via geodesic distances to represent the lower bound that guarantees a correct solution.
Empirical evidence linking geodesics and common neighbors is, however, still lacking~\cite{burch2020state}. 

A task instance complexity measure for Task~1 should~therefore account for: 1)~the closeness of $w$ to $u$ and $v$, and 2)~far $w$
which may be easier to over\-look.
Let us construct a sphere $S_{uv}$ around $u$ and $v$, with center $o$ and radius the distance $\frac{||u - v||}{2}$.
If $w$ is inside $S_{uv}$, $w$ is geometrically between $u$ and $v$ and it falls in sight while looking at both:~finding $w$ is easy.
If $w$ is outside $S_{uv}$, additional (or longer) searches are needed to find $w$ and its incident edges may not be easily traceable.
For the signal instance complexity of Task~1, we propose:
\begin{equation}\label{TIC_CN_S}\tag{1}
    \xi_1(\Gamma) = \sum_{w \in N_{uv}} \Bigl(dist(w,u) + dist(w, v) - dist(u,v)\Bigr)^2
\end{equation}

The difference between the definition above and the simple sum of the distances of $w$ to $o$ is easy to see when $w$ belongs to segment $\overline{uv}$. 
In the simple sum, a node $w$ on $\overline{uv}$ would be awarded a higher cost, despite being very close to $u$ or $v$. 
In Eq.~\ref{TIC_CN_S}, all nodes $w$ on $\overline{uv}$ have a value of zero.
Now by taking a low polynomial, we punish nodes that are beyond $S_{uv}$, though less strongly.

Increases in visual clutter are known to hinder understanding and performance~\cite{purchase1997experimental, purchase1997aesthetic, ware2002cognitive, huang2007using}.
For noise instance complexity, the ROI should include any $w$, as well as $u$ and $v$.
For the first strategy~(with $u$, the lower-degree node), the ROI should also include any neighbor of $u$, as each neighbor of $u$ has to be inspected at least once.
To represent the ROI, we construct a sphere $S$, with center $o$ and radius the distance $dist(o, far(u))$ from $o$ to the furthest neighbor of $u$.
The sphere $S$ serves as a simple and uniform estimator of the view---solving the task requires looking within $S$.
We estimate the noise in the ROI based on the length $\mu(x)$ of all graph elements $x \in S$ not part of the solution:
\begin{equation}\label{TIC_CN_N}\tag{2}
    \noise{\xi_1}(\Gamma) = \sum_{\substack{x \not\in N_{uv}\cup E_{uv}\\\text{and } x \in S}} \mu (x)
\end{equation}

\subsubsection{Task Instance Complexity of Task 2}\label{task2}
In determining the shortest path between two nodes, users tend to prioritize search in the direction of the target \mbox{node~\cite{ware2002cognitive, huang2009graph}}.
Starting from the source node, incident edges are added to the path, while changes in closeness to the target node are used to track progress.
During this process, users perform a series of quick scans along the line of sight to the target node.
Known as the geodesic-path tendency~\cite{huang2009graph}, this has been observed, e.\,g., in the reading of public transit maps~\cite{schafer2022group}.
Short paths tend~to~be approximated altogether~(path-continuity tendency)~\cite{ware2002cognitive, burch2020state}.
There seems to be a strong tendency to map topological onto spatial distances under the constraint of minimality.
Edges that go away from the target node are thus less likely to be searched.
These necessarily form sharp node angles that have been found to adversely affect graph comprehension~\cite{huang2007using, huang2014larger}.

Let $u...v$ be the shortest path such that $deg(u) \leq deg(v)$. 
Let $\theta~=~\sphericalangle \left(prev(w) - w, next(w) - w\right)$ be the node angle at $w$ for $prev(w),\ w,\ next(w) \in u...v$.
When $prev(w)$ or $next(w)$ do not exist, $\theta$ is $0$.
Note, $\theta$ is always the smaller of the two angles.
The signal instance complexity of finding the shortest path is computed as:
\begin{equation}\label{TIC_SP_S}\tag{3}
    \xi_2(\Gamma) = \max_{\forall u...v}\left\{\frac{1}{2} \sum_{w \in u...v} \left(1 - cos(180^\circ - \theta)\right) \cdot dist(w, v)\right\}
\end{equation}

In Eq.~\ref{TIC_SP_S}, the geodesic paths $dist(w, v)$ are weighted by the angle difference for which there are three cases: 

\begin{itemize}
    \item[-] $\theta > 90^\circ$: the polyline $(prev(w), w), (w, next(w))$ approximates a line or a curve. The user can more easily follow the outgoing edge $(w, next(w))$ and $dist(w, v)$ is adjusted with a lower weight.
    \item[-] $\theta \approx 90^\circ$: the outgoing edge $(w, next(w))$ is near-orthogonal to the ingoing edge $(prev(w), w)$. The user is not approaching $v$ or zigzagging toward $v$ and the weight of $dist(w,v)$ is about $1$.
    \item[-] $\theta < 90^\circ$: the polyline $(prev(w), w), (w, next(w))$ approximates a sharp bend. The user has to search more for the outgoing edge $(w, next(w))$ and $dist(w, v)$ is adjusted with a higher weight.
\end{itemize}

With $\frac{1}{2}$ we normalize the range of \mbox{$1 - cos(180^\circ - \theta)$}.
In case of several shortest paths $u...v$, the $\max(\cdot)$ in Eq.~\ref{TIC_SP_S} gives the lower bound~(see Sec.~\ref{defs}).
The measure differentiates paths of different shapes and it scales well with the number of nodes.
Let us fix the node positions in space and insert nodes in the shortest path. 
Any inserted node $w'$ is a common neighbor of two other nodes and the previous edge $(prev(w'), next(w'))$ is replaced by two new edges.
Then:
\begin{itemize}
    \item[-] $w'$ is on or near the previous edge: the geodesic distances are adjusted for no or little increase in complexity.
    \item[-] $w'$ is away from the previous edge: $w'$ deviates from the geodesic-path tendency and the geodesic distances are adjusted for a larger increase in complexity.
\end{itemize}

In Eq.~\ref{TIC_SP_S}, the adjusted distance is in absolute terms.
Instead, one can set $dist(u, v) = 1$ to make $dist(w,v)$ relative after which the measure becomes dimensionless and scale-invariant.

\begin{figure}[!t]
    \centering
    \includegraphics[width=\linewidth]{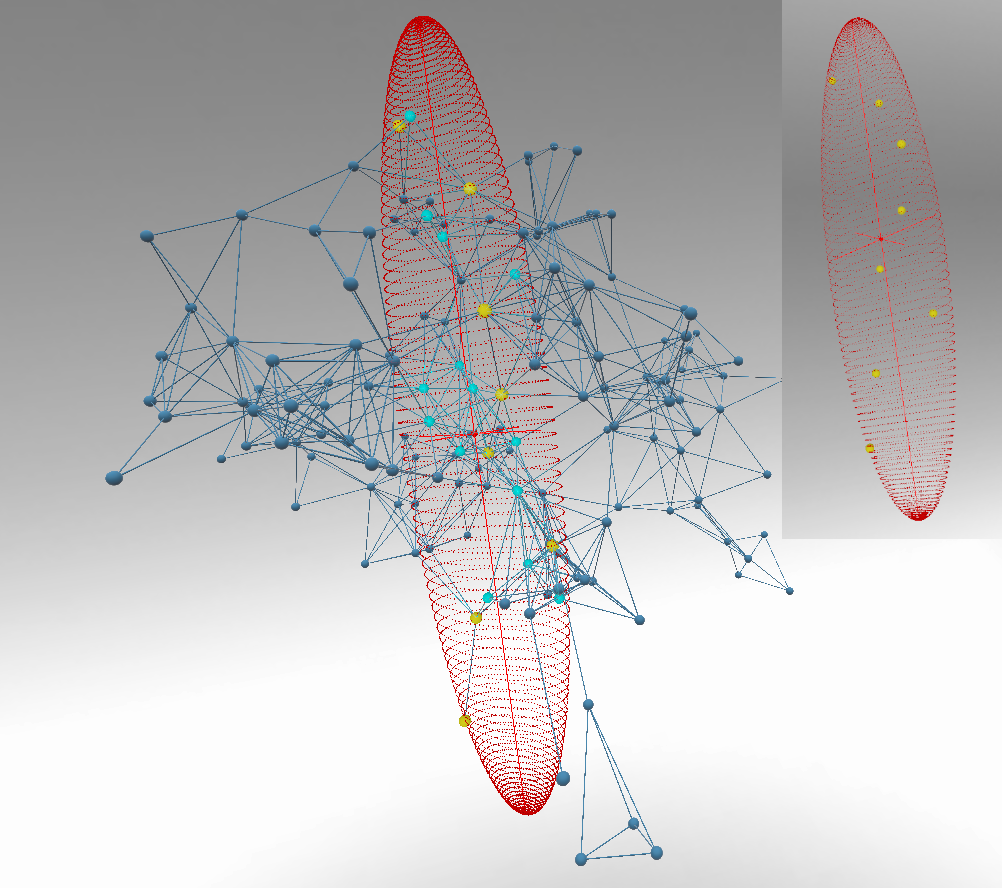}
    \caption{Path nodes $w\in u...v$~(yellow) are enclosed by their minimum volume ellipsoid $E_3$~(red), as seen in the small view. Nodes and edges not part of $u...v$ that lie wholly or partly in $E_3$ are then counted as noise~(cyan).}
    \label{fig:SP_N}%
\end{figure}%
For a given geometric container~\cite{wagner2005geometric} enclosing the path in $\Gamma(G)$, we estimate the noise by how one looks at the path.
To determine a suitable geometric container, we overlay all path instances.
We transform each path $u...v$, so that all $u$ have the same start and all $v$ have the same end position in $\Gamma(G)$; geodesic path distance set to $1$.
Based on the shape of all instances~(see Suppl. Mat.), an ellipsoid is inferred as the geometric container.
For each task instance, we find the minimum-volume enclosing ellipsoid~\cite{moshtagh2005minimum}, where the points used are the positions of the nodes $w \in u...v$~(see Fig.~\ref{fig:SP_N}).
Any node or edge not part of the path $u...v$ is counted as noise,
when it intersects the interior of the ellipsoid.
Compared to a sphere, the ellipsoid approximates $u...v$'s shape more closely once the radius constraint is relaxed.
It shows how one looks at the path in 3D: enclosed elements which may interfere with path-following are visible from different perspectives.
The noise is defined in Eq.~\ref{TIC_SP_N} where the measure $\mu(x)$ of an element $x \in G$ is its length.
For several shortest paths $u...v$, the $\min(\cdot)$ gives the lower bound on the noise~(see Sec.~\ref{defs}).

\begin{equation}\label{TIC_SP_N}\tag{4}
    \noise{\xi_2}(\Gamma) = \min_{\forall u...v}\left\{\sum_{\substack{x \not\in u...v\text{ and}\\ x \in E_3}} \mu (x)\right\}%
\end{equation}%

\subsection{Hypotheses}

\begin{enumerate}
    \item [(\textbf{H1})] An ad hoc pair has higher accuracy than an individual and a nominal pair.
\end{enumerate}
A higher accuracy for the ad hoc pairs is expected in spite of process loss~\cite{steiner1972group, hackman2010group}.
For higher common-neighbor counts and node degrees~(Task\,1), memorization could outdemand identification~\cite{wang2006layered}~(and benefit collaboration), whereas differences in accuracy should be less apparent for lower counts~\cite[Exp.\,1]{prouzeau2016evaluating}.
For longer and denser paths~(Task\,2), the ad hoc pairs should achieve higher accuracy also, e.g., via error correction~\cite{thorndike1938effect, kameda2022information} and the use of multiple perspectives~\cite{sando2011impact}. 

The difference to the nominal pairs should thus be only smaller, confirming the presence of collaborative~\mbox{effects}.
We hypothesize that collaborative effects may emerge from interaction in the pair~\cite{hackman1975group, schwartz1995emergence} beyond just cognitive resources, which are held comparable~(via the \mbox{nominal} pair).
The effects may be caused by the application of appropriate strategies~\cite{schwartz1995emergence}, iterative refinements of understanding, and the synchronization of cognitive~\mbox{processes}~\cite{hackman2010group, clark1991grounding}, such as perception and memory.
Besides verbal communication, the visuospatial aspects of the CVE, e.\,g., engagement~\cite{marriott2018immersive}, shared awareness~\cite{kiyokawa2002communication, cordeil2016immersive}, and nuanced pointing~\cite{heer2007design, billinghurst2018collaborative}, are also expected to support collaboration, so that the ad hoc pairs can have higher accuracy than the nominal pairs.

\begin{enumerate}
    \item [(\textbf{H2})] An ad hoc pair has slower completion time than an individual, but is faster than a nominal pair.
\end{enumerate}

The tasks we pose are of mixed composition~(see Sec.~\ref{defs}), and the workload is divisible.
If we apply Amdahl's law~\cite{gustafson1988reevaluating}, the theoretical upper speedup of a pair is two, or twice as fast.
This is, however, unlikely given that the ad hoc pairs will need to actively communicate and coordinate more~\cite{dourish1992awareness}, i.e., verbally, referentially, and spatially.
Reaching consensus, searching overlooked neighbors, and quickly identifying shorter paths will be less present in or executed more efficiently by the individuals.

Then, the question arises: does collaborative solving reduce the time difference of two individuals solving the task separately?
We have graph analytical tasks of varied instance complexity in an environment that should predispose information pooling~\cite{hill1982group, kameda2022information}, coordinated use of perspectives~\cite[Fig.\,9]{lee2020shared}, and the emergence of trust~\cite{heer2007design} to allow temporal gains from collaboration.
For the setting under test, we expect the ad hoc pairs to be faster than the nominal pairs.

\begin{enumerate}
    \item [(\textbf{H3})] With increased task instance complexity, 1)~the difference in accuracy increases, that is between ad hoc pair vs. indi-vidual and nominal pair; and 2)~the difference in completion time decreases between ad hoc pair and individual, and increases between ad hoc pair and nominal pair.
\end{enumerate}

We expect task instance complexity to interact with group type.
The interaction would confirm the utility of collaborative solving when instance complexity changes.
As a factor for group outperformance first hypothesized about 50 years ago~\cite{shaw1976assembly}, task complexity, across its many definitions, has remained empirically elusive~\cite{hill1982group, kerr2004group, laughlin2011group}.
Motivated by this new and untested integration of environment and visuospatial tasks, we think collaborating pairs can pool their resources and tackle higher instance complexities better than separate individuals.
About accuracy and time, we expect the differences in \textbf{H1} and \textbf{H2} to scale with task instance complexity. 

When task instance complexity is increased, searching and identification of the necessary interactions with graph elements will be superior and better parallelized in the ad hoc pairs.
Unlike the individuals and nominal pairs, the ad hoc pairs are expected to communicate relevant aspects of the task instances to reach a consensus.
Error correcting~\cite{thorndike1938effect, kameda2022information} will further support the ad hoc pairs, at least accuracy-wise, and more so on Task~2.
Since hand and gaze interactions are excluded, division of workload will not be that evident and additional coordination and strategy building will be required.
Multiple perspectives will be found more effectively~\cite{sando2011impact} and shared in the ad hoc pair~\cite{poretski2021physicality}, thus reducing the effects of visual clutter.
Concurrently, consensus reaching from interaction in the ad hoc pair~\cite{sherif1935study, dyer2009leadership} should account for savings in completion time.
As a result, the ad hoc pair is expected to become progressively more accurate and faster against the nominal pair and the individual as task instance complexity increases.

\subsection{Additional observation}
Measuring and analyzing cognitive load in its entirety would warrant a study of its own.
Instead, we incorporate a previously operationalized subjective measure of cognitive workload~\cite{kosch2023survey} in our study design.
The investigation into \textbf{{\scriptsize RQ}{\small 3}} is observational and across all task instances.

\begin{enumerate}
    \item [(\textbf{O1})] Total task instance complexity affects the cognitive workload of ad hoc and nominal pair members differently.
\end{enumerate}

Prolonged cognitive work causes mental fatigue from meta\-bolic regulation in the brain~\cite{wiehler2022neuro}.
Collaborating in CVE can mean varying cognitive workload~\cite{bai2020user, tong2023towards}.
On the one hand, division of workload, insights generation, and pooling of~information may decrease the impact of task instance complexity.
On the other hand, spatial and temporal coordination or uneven division of workload may increase it.
We expect collaboration to interact with task instance complexity, but cannot set the direction.

\subsection{Experimental procedure}~\label{procedure}

We ran a pilot study with $8$ participants across two locations based on which we could test and reiterate the experimental procedure, the software prototype, and the study documents.
We built our prototype on four Meta~Quest Pro with video pass-through and inside-out tracking~(v62)---a viable option at the time of the study.
In both locations, each pair of headsets was connected to a dedicated server in a local area network.

Participants could determine the language of choice themselves~(IT, DE, or EN). 
After a short introduction, participants signed an informed consent agreeing to participate in the study.
Several covariates were considered, using a demographics and prior-experience questionnaire.
Participants were introduced to each of the three main topics.
As part of learning about graphs, participants were asked to solve three training tasks on paper and in a casual setting---ad hoc pairs could also talk to each other.
Participants were instructed how to put on the headset, where to start, what the general format is, and how to adjust eye focus if needed.
Participants were then walked through an illustrative example involving a book rendering: the goal was to get familiar with viewing, moving, and typing, as well as to ask any questions.
After completing all task instances, participants were asked to fill out NASA-TLX and participate in a short interview, before final debriefing.
A procedure lasted between 1 and 1.5 hours, with few ad hoc pairs around the 2-hour mark.

Following the controlled randomization, the correct answers varied from $2$--$11$ for Task~1, and $3$--$16$ for Task~2; distributions were positively skewed and had heavy tail~(see Suppl. Mat.).
We logged data, such as accuracy and completion time, using the headset.
Due to the high temporal resolution, small time differences appeared in the ad hoc pairs for which the longer of the two completion times was used.
The first instance in each series was discarded, as well as any other instances where interruptions occurred~($4$), plus one faulty instance.

\subsection{Participants}
$72$ participants took part in the study which was evenly split in two countries:~Italy and Germany, and conducted in three languages~(EN, DE, IT).
Mean age was $25.3$ years ($SD = 7.6$, range:~$19-59$), with $35$ female and $37$ male.
Education level included bachelor~($39$), master~($20$), doctoral~($9$), or other~($4$), across $18$ fields.
Except $5$, all participants reported normal or normal-to-corrected vision.
The small number of participants with abnormal vision was kept to promote ecological validity.
$41$ were familiar with networks, $25$ with visualizations, and $25$ with groups.
Due to controlling and extraneous circumstances we discarded and redid sessions with $15$ persons~($10$ together, $5$ alone). 
No participants withdrew from the experiment.

\section{Results}

\begin{figure*}[t]
    \centering
    \subfloat[\hspace*{-1.25cm}]{{\includegraphics[width=.67\columnwidth]{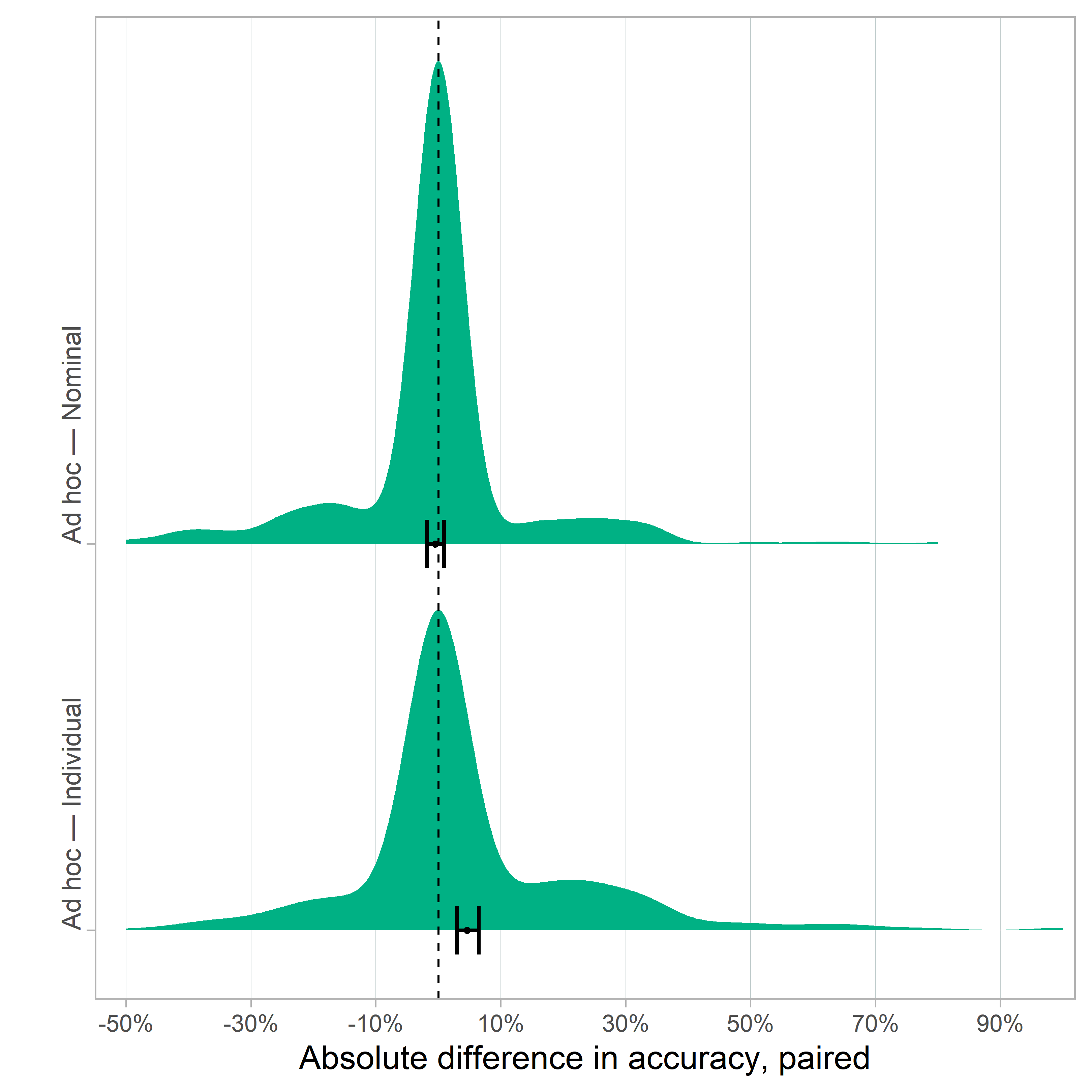} }}%
    \subfloat[\hspace*{-0.5cm}]{{\includegraphics[width=.67\columnwidth]{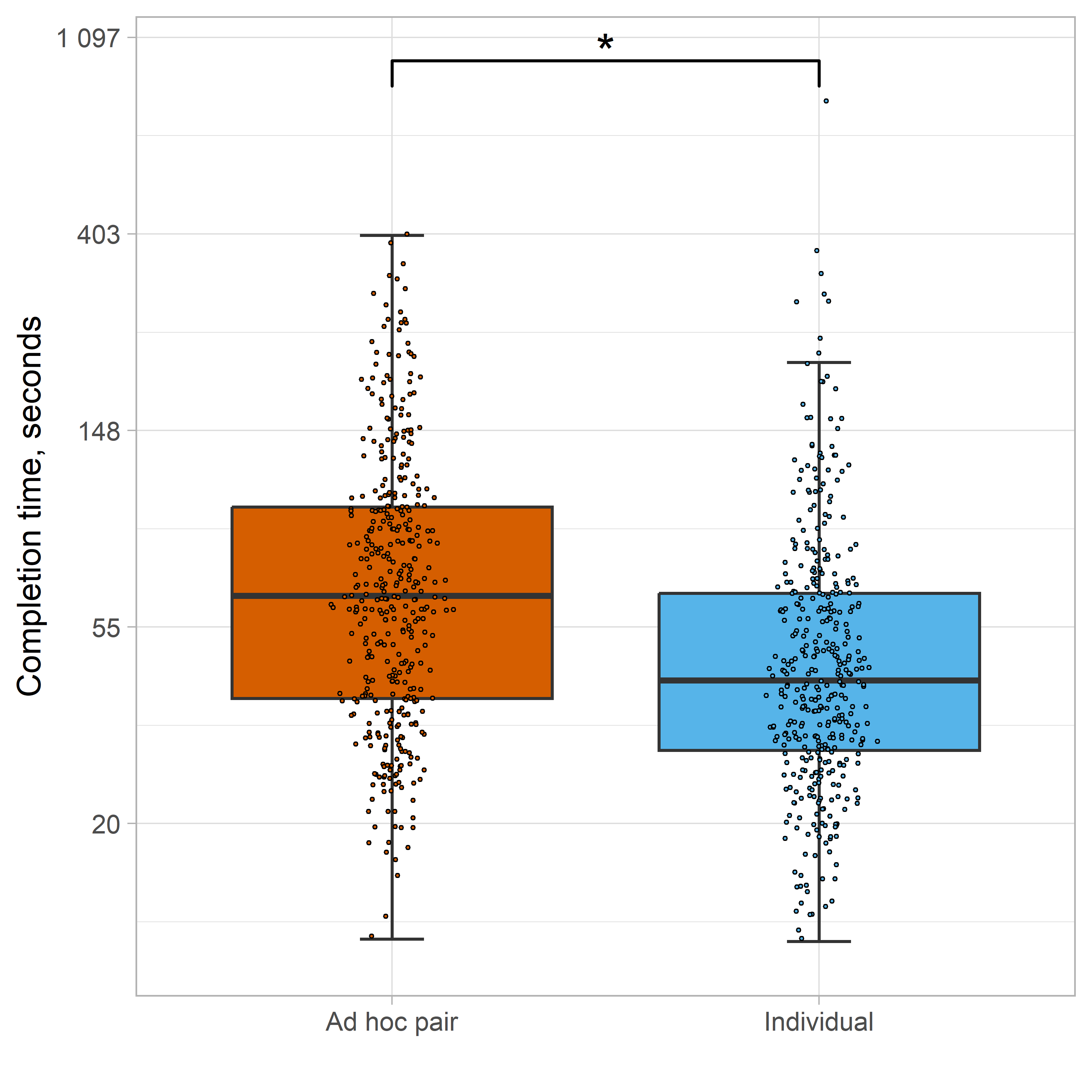} }}%
    \subfloat[\hspace*{-0.5cm}]{{\includegraphics[width=.67\columnwidth]{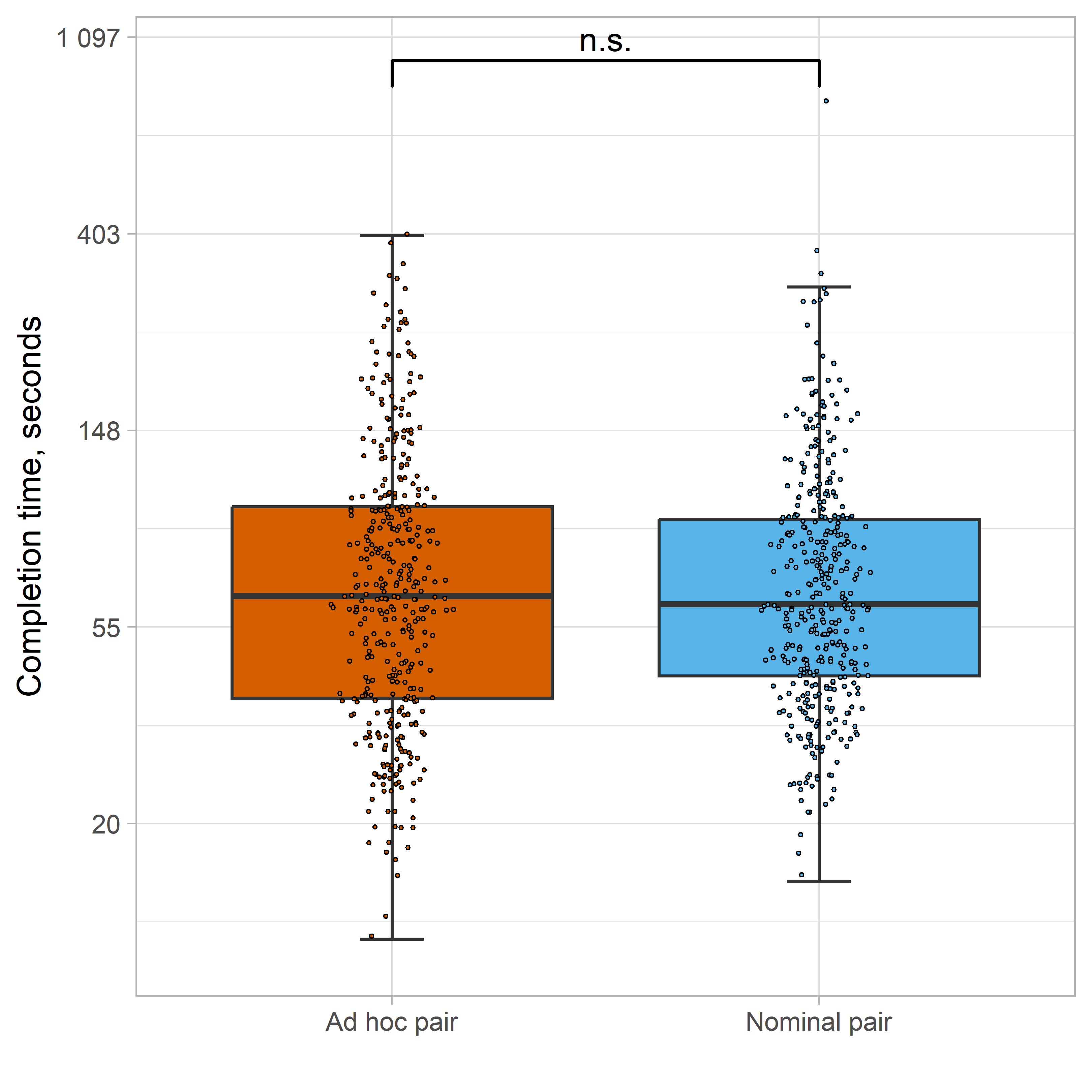} }}%
    \caption{
    a)~Absolute difference in accuracy, paired, between group types.
    Error bars mark significant difference from the null, based on bootstrapped $95$\% CI.
    b), c)~Average completion time, seconds, given by group type.
    Horizontal bars~(top) indicate significance after post hoc correction.
    }%
    \label{fig:acc_completiontime}%
\end{figure*}

\subsection{Quantitative results}
The starting point of our \mbox{analysis} was a linear mixed-effects model~(Suppl.\,Mat., Eq.~S2) and \mbox{alpha} level of $0.05$.
We considered mixed-effects modeling to account for individual variability and within-subject correlation in estimating population averages~\cite{meier2022anova}.
We also used stratified bootstrap resampling where necessary instead of just reporting p-values~\cite{aichem2024emphasise}, e.\,g., non-parametric rank tests.
All statistical models, assumptions, and plots used in verifying these assumptions are in the Suppl. Mat, published together with all scripts~\cite{darus-4231_2024}.

\subsubsection{\textbf{Accuracy}~(\textbf{H1})}
No transformation could meet the assumptions of our linear model.
We applied non-parametric bootstrap resampling~(\mbox{\small $R=10^4$}) to obtain estimates for the paired mean difference.
Resampling was stratified to account for repeated measures.  
Bootstrap resampling guarantees that the sampling distribution converges to a normal distribution.
A paired permutation test was used to obtain p-values.

\paragraph{Ad hoc pair---Individual}
An ad hoc pair's average accuracy was $93.9\%$, while an individual could achieve $89.2\%$.
Results~(Fig.~\ref{fig:acc_completiontime}~(a), bottom) evidenced the ad hoc pair as having a $4.6\%$ higher mean accuracy with $95$\%~CI~$\left[2.9\text{\%}, 6.5\text{\%}\right]$. 
The paired permutation test p-value stood at $p=0.0001$.

\paragraph{Ad hoc pair---Nominal pair}
The average accuracy of  a nominal pair was $94.4\%$.
Results~(Fig.~\ref{fig:acc_completiontime}~(a), top) lacked evidence of a clear advantage: $0.5\%$ in favor of the nominal pair with $95$\%~CI~$\left[1.9\text{\%}, -0.9\text{\%}\right]$. 
The paired permutation test found no significant difference.

\subsubsection{\textbf{Completion time}~(\textbf{H2})}
We applied linear mixed-effects modeling~(Suppl.\,Mat.,\,Eq.\,S2).
After logarithmic transformation, residuals were log-normally distributed and homoscedastic.
Random effects appeared normally distributed despite~the smaller sample size in each group~($36$).
For increased robustness and better $95$\%~CI estimates, we used stratified bootstrap resampling~(\mbox{\small $R = 10^4$}) refitting each model.

\paragraph{Ad hoc pair---Individual}
The average measured completion time of an ad hoc pair was $83.17s\pm64.74$, whereas the average individual used $57.59s\pm59.87$.
The geometric mean completion time
was $65.16s$ and $44.73s$, respectively.
The two times were significantly different~({\small $t=-3.032$, $df=34$, \mbox{$p=0.00463$}}), estimating the ad hoc pair's completion time to be \mbox{$e^{0.37628} \approx 1.46$} times higher~(Fig.~\ref{fig:acc_completiontime}~(b)). 
The bootstrapped $95$\% CI were $(61.31s$, $69.27s)$ for the ad hoc pair, and $(38.91s$, $51.46s)$ for the individual, estimating an ad hoc pair to be between $1.58$ and $1.35$ times slower. 

\paragraph{Ad hoc pair---Nominal pair}
The average measured completion time of a nominal pair was $80.00s\pm64.35$, compared to the ad hoc pair's $83.17s\pm64.74$.
The nominal pair's geometric mean completion time was $65.50s$.
There was no significant difference between the two means~(Fig.~\ref{fig:acc_completiontime}~(c)), which was consistent with bootstrapping.

\subsubsection{\textbf{Task instance complexity,~\texorpdfstring{$\xi$}{Lg}},~(\textbf{H3})}
We treated $\xi$ as fixed and not a random effect, since the graphs which task instances were randomly drawn from had been selected of a specific type~(see Sec.~\ref{netvars}).
Interaction between $\xi$'s components was excluded.
Noise instance complexity followed a log-normal distribution and was transformed accordingly.
The two components were not correlated.

\paragraph{Accuracy~(\textbf{H3},~1))}
Accuracy was bounded in $\left[0, 1\right]$ and could not follow a normal distribution.
A logistic transformation
was not applicable due to a bimodal distribution of the residuals.
Mixed-effects logistic regression by means of a generalized linear mixed-effects model~\cite{dixon2008models} was instead used to model the probability of an accurate answer~(Suppl.\,Mat., Eq.~S4).
Our analysis revealed no clear violations of the model's assumptions, as
incorporating curvilinear components supported the log-transformation of noise instance complexity.

A Wald test found no specific evidence toward improved ad hoc pair's odds as task instance complexity increased~($\beta_3$ coefficient in Suppl. Mat., Eq.~S4), which was the case for both the ad hoc---individual and the ad hoc---nominal comparisons.

\paragraph{Completion time~(\textbf{H3},~2))}
We extended our model from \textbf{H2} to include $\xi$~(see Suppl. Mat., Eq.~S5).
After log-transformation of completion time, residuals were normally distributed and homoscedastic in each of the two models,  
while random effects appeared normally distributed.

\par\indent\indent\textit{b.1) Ad hoc pair---Individual.} 
For increased task instance complexity, the difference in completion time between the ad hoc pair and the individual did not decrease.
In fact, the ad hoc pair became slightly slower, but not significantly.

\par
\begin{figure}[t]
    \centering
    \subfloat{{\includegraphics[width=0.24\textwidth]{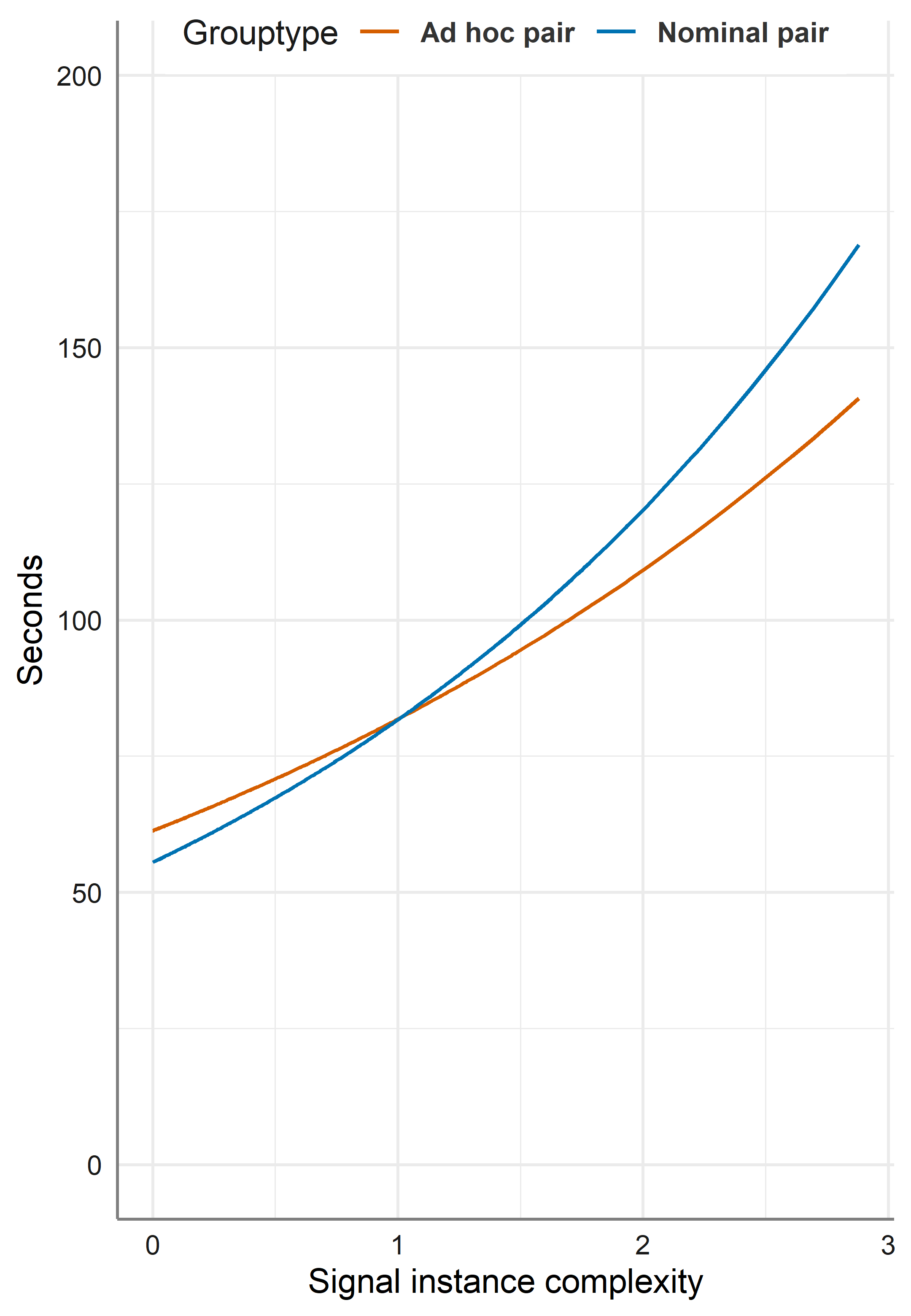} }}%
    \subfloat{{\includegraphics[width=0.24\textwidth]{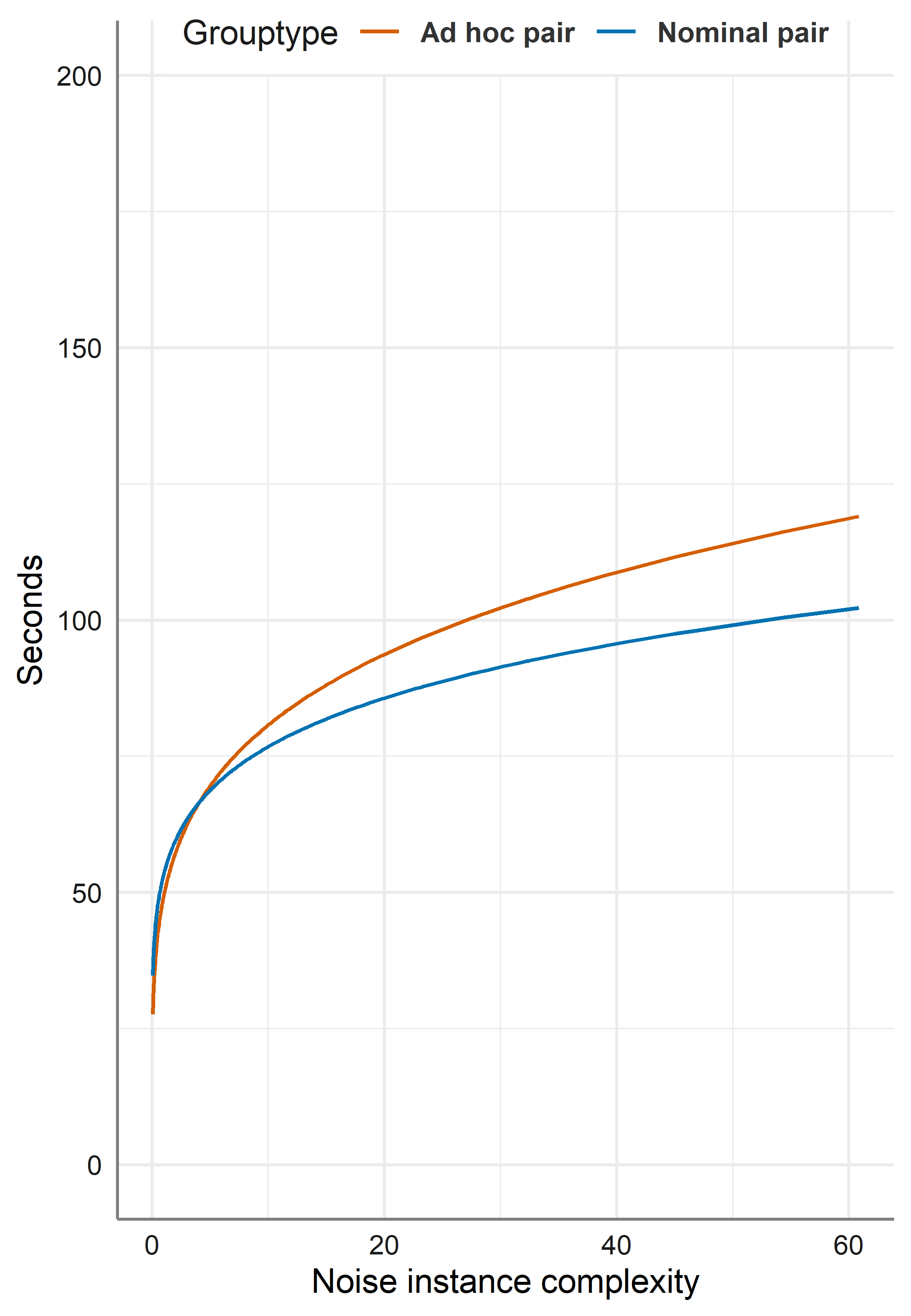} }}%
    \caption{Completion time by task instance complexity for ad hoc and nominal pairs.
    As signal instance complexity increases~(left), nominal pairs tend to slow down more than ad hoc pairs.
    For increases in noise instance complexity~(right), the difference between pair types remains largely unchanged.}%
    \label{fig:ct_H3_AN}%
\end{figure}

\indent\indent\textit{b.2) Ad hoc pair---Nominal pair.} 
The difference in completion time between the ad hoc pair and the nominal pair increased with task instance complexity~(Fig.~\ref{fig:ct_H3_AN}).
Our bootstrapped model suggested a small evidence of interaction with signal instance complexity.
As this increased, the nominal pair's completion time increased $e^{0.098} \approx 1.1$~times proportionally more~(bootstrapped $95$\%~CI~$\left[1.01, 1.21\right]$).
The evidence of interaction with noise instance complexity was inconclusive, 
though marginally in favor of the nominal pair.

\subsubsection{\textbf{Cognitive workload},\,CW,~(\textbf{O1})}
NASA-TLX was administered after the completion of all task instances.
We followed Hart's approach~\cite{hart1986nasa} and computed the overall weighted rating for each participant.
Responses ranged between $0$~(low) and $100$~(high), where exact totals of $0$ and $100$ are unlikely.
Responses approximated a beta distribution 
based on which we fit a beta regression model~\cite{ferrari2004beta} of the probability to have high CW for the combined task instance complexity~(see Suppl. Mat., Eq.~S6).
To reduce multicollinearity, we centered $\xi$'s components around their mean.
Plotting of deviance residuals showed no particular deviations.
Two responses corresponding to levels of low CW~($17.00$, $23.67$) were flagged as outliers, which we kept after closer inspection, e.\,g., when $\xi$'s components crossed. 

\begin{figure}[b]
    \centering
    {\includegraphics[width=\columnwidth]{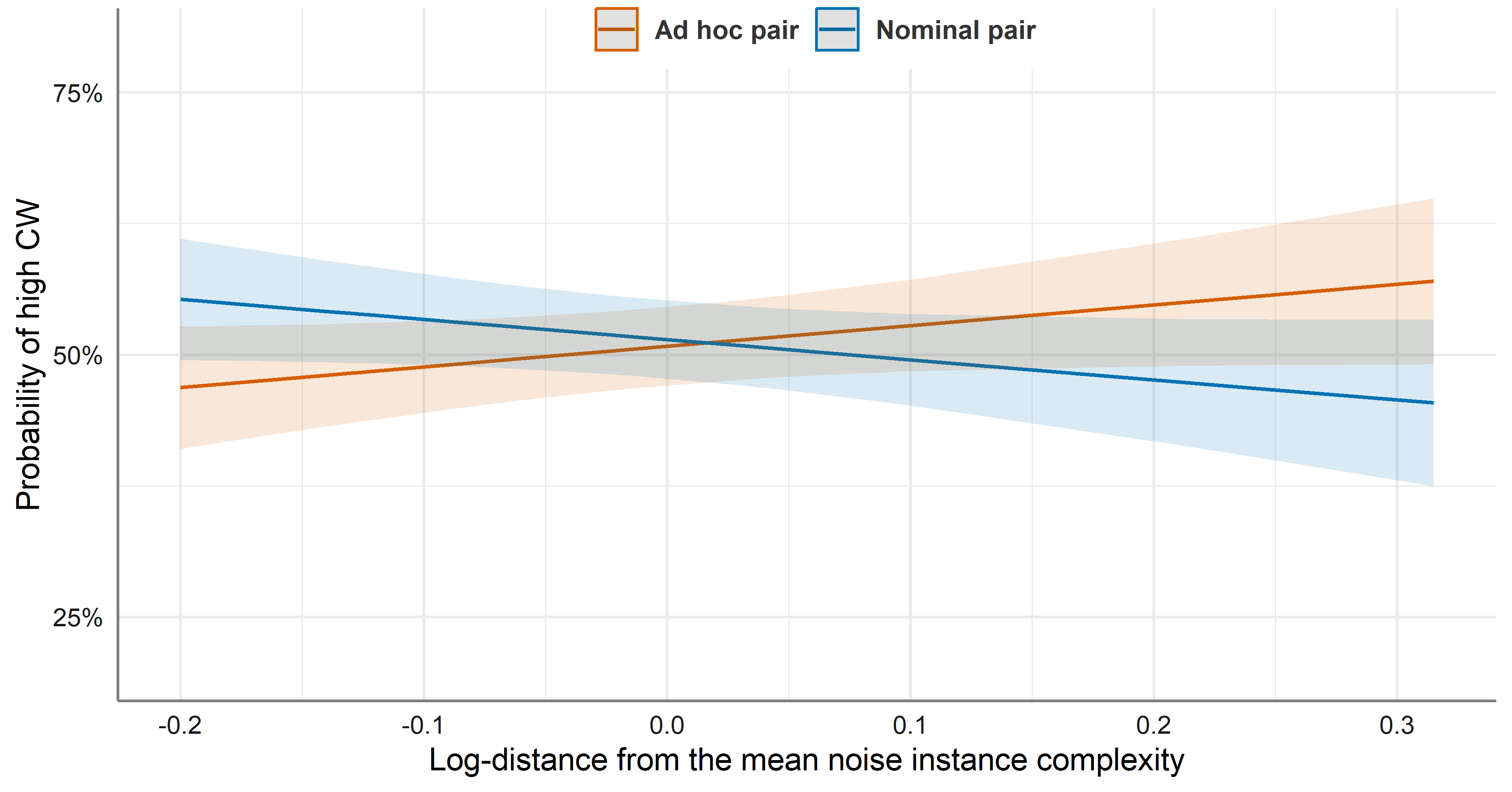} } 
    \vspace*{-0.5cm} 
    \caption{Average probability of high cognitive workload~(CW) for total noise instance complexity from the sample mean of $194m$.
    }%
    \label{fig:CW_H4_Noise}%
\end{figure}

\textit{a) Ad hoc pair---Nominal pair.}
When total noise instance complexity increased, the ad hoc pair's probability of high CW significantly increased~({\small $z = 2.397$, $df = 7$, \mbox{$p = 0.0165$}}); when it decreased, the opposite was also true~(Fig.~\ref{fig:CW_H4_Noise}).
The effect was pronounced: 
the odds of high CW were predicted to change by $15\%$ for every $e^\text{0.1}$ change in total noise instance complexity from the sample mean of $194m$~($95$\%~CI $\left[3\text{\%}, 25\text{\%}\right]$).
The interaction with the total signal instance complexity was not statistically significant.

\subsubsection{\textbf{Post hoc analysis}}
Following a recent discussion on correcting for multiple testing in applied research~\cite{garcia2023use}, we divided inferences into three families:~1)~ad hoc \mbox{pair---individual} performance; 2)~ad hoc---nominal performance; and 3) effects of paired solving on cognitive workload for combined task instance complexity~(\textbf{O1}). 
These account for the nesting of tests in a hypothesis and consider the atomicity of responses, such as accuracy and time.
As nominal pairs cannot be compared with individuals in our study design, an omnibus null hypothesis cannot be tested and correction is not necessary.
To control the error rate of false-positive inferences in each family, we used the Holm-Bonferroni correction.
Adjusted \mbox{p-values} did not markedly differ from reported p-values, see Suppl. Mat.~\cite{darus-4231_2024}.

Our previous analysis covered both tasks collectively.
Therefore, we re-examined results post hoc and found these to be consistent across Task~1 and Task~2.
In the following, we report on how task instance complexity affects responses independent of group type.

\paragraph{Accuracy}
The effect of noise instance complexity on accuracy was significant~({\small $z=-5.514$, $p<0.0001$; 
$z=-5.518$, \mbox{$p<0.0001$}}); that of signal instance complexity was not. 
\nolinebreak
The odds of an accurate answer were predicted to decrease by $49\%$ for every $e^1$ increase in noise instance complexity~($95$\%~CI $\left[35\text{\%}, 60\text{\%}\right]$).
Fig.~\ref{fig:marginal_effects_plots_H3} shows how the average predicted probability of an accurate answer decreases as noise instance complexity increases.

\paragraph{Completion time}
Signal instance complexity had a significant effect~({\small $t=7.301$, $df=753.1$, \mbox{$p < 0.0001$}; 
$t=7.463$, $df=753.4$, $p<0.0001$}):~
completion times were predicted to increase by $33\%$~(bootstrapped $95$\%~CI $\left[25\text{\%},43\text{\%}\right]$) for \mbox{every} increase in \mbox{signal} \mbox{instance} complexity.
Noise instance complexity also had a significant effect~({\small $t=10.202$, $df=753.8$, \mbox{$p<0.0001$}; 
\mbox{$t=10.426$}, $df=754.3$, \mbox{$p<0.0001$}}): 
completion times were predicted to increase by \mbox{$24\%$}~(bootstrapped $95$\%~CI $\left[18\text{\%},30\text{\%}\right]$) for each $e^1$ increase in noise instance complexity.

\paragraph{Cognitive workload}
Total signal instance complexity had a small significant effect~({\small $z = -2.219$, $df = 7$, \mbox{$p = 0.0265$}}).
For every increase from the sample mean of $16.5$, the odds of high CW were predicted to decrease by $11$\%~($95$\% CI $\left[1\text{\%}, 20\text{\%}\right]$)
Total noise instance complexity had larger but not significant effect on increasing the odds of high CW. 

\subsection{Qualitative results}
In the following, we report on the interview where we asked participants about subjective and difficult to quantify aspects.

\subsubsection{\textbf{Perceived task difficulty}}
Each participant rated task difficulty a on scale of one~(the least) to five~(the most) and explained the choice in their own words.
Ad hoc pairs found on average each of the two tasks slightly easier than nominal pairs (Task~1:~$2.66$~{\small$\xleftrightarrow{\text{vs.}}$}~$2.88$, Task~2:~$3.08$~{\small$\xleftrightarrow{\text{vs.}}$}~$3.22$), with the shortest path task~(SP) slightly more difficult than the common neighbors task~(CN).
By frequency, participants' explanations why SP was more difficult included: a larger solving space physically and analytically, an increased sense of uncertainty, and being a more visually demanding task, e.\,g., due to density.
For CN, feedback pointed to greater memory involvement and being a less intuitive task.
SP and CN thus made for a more global and a more local systematic tasks, respectively. 

\begin{figure}[t]
    \centering
    \subfloat{{\includegraphics[width=0.24\textwidth]{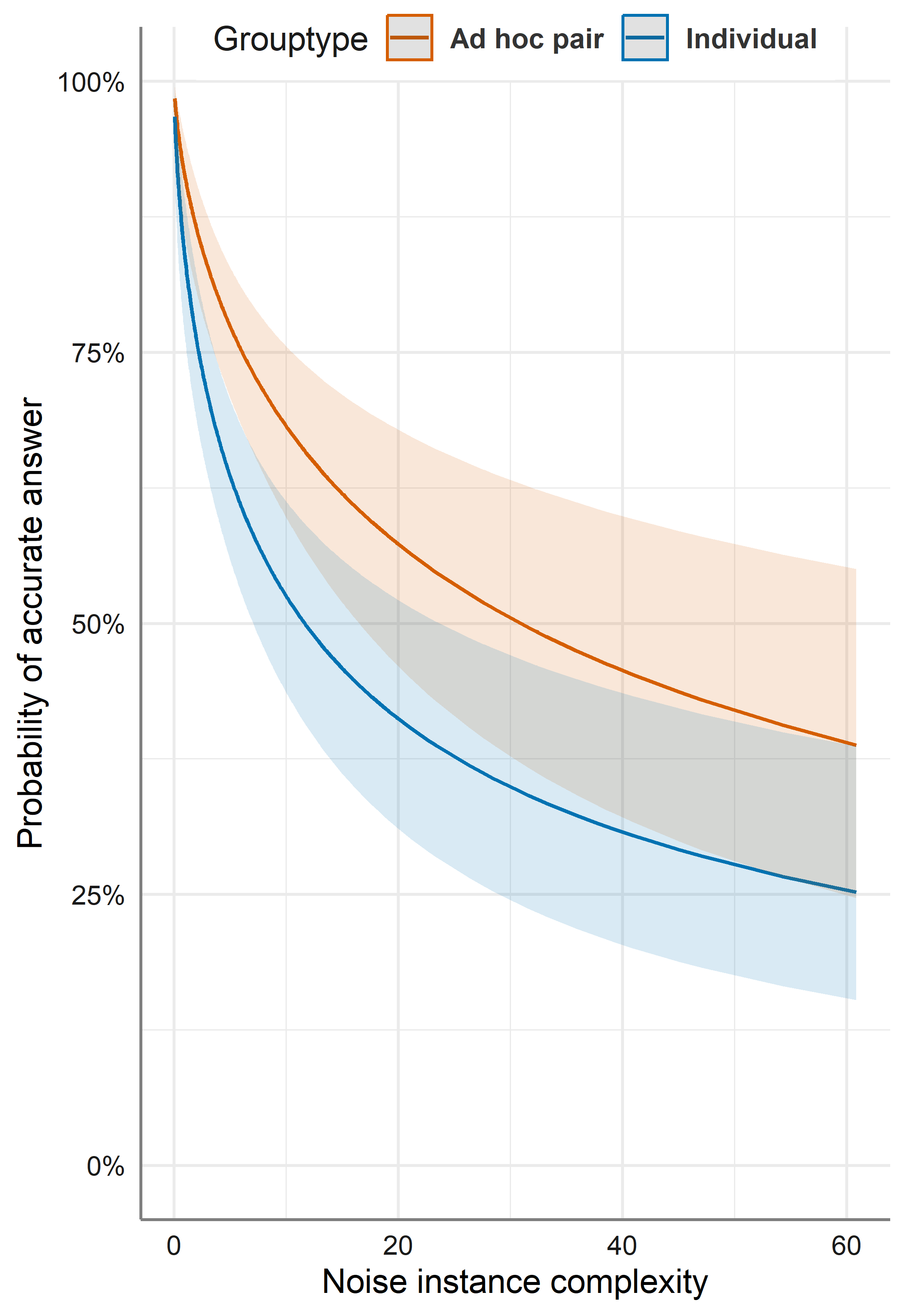} }}%
    \subfloat{{\includegraphics[width=0.24\textwidth]{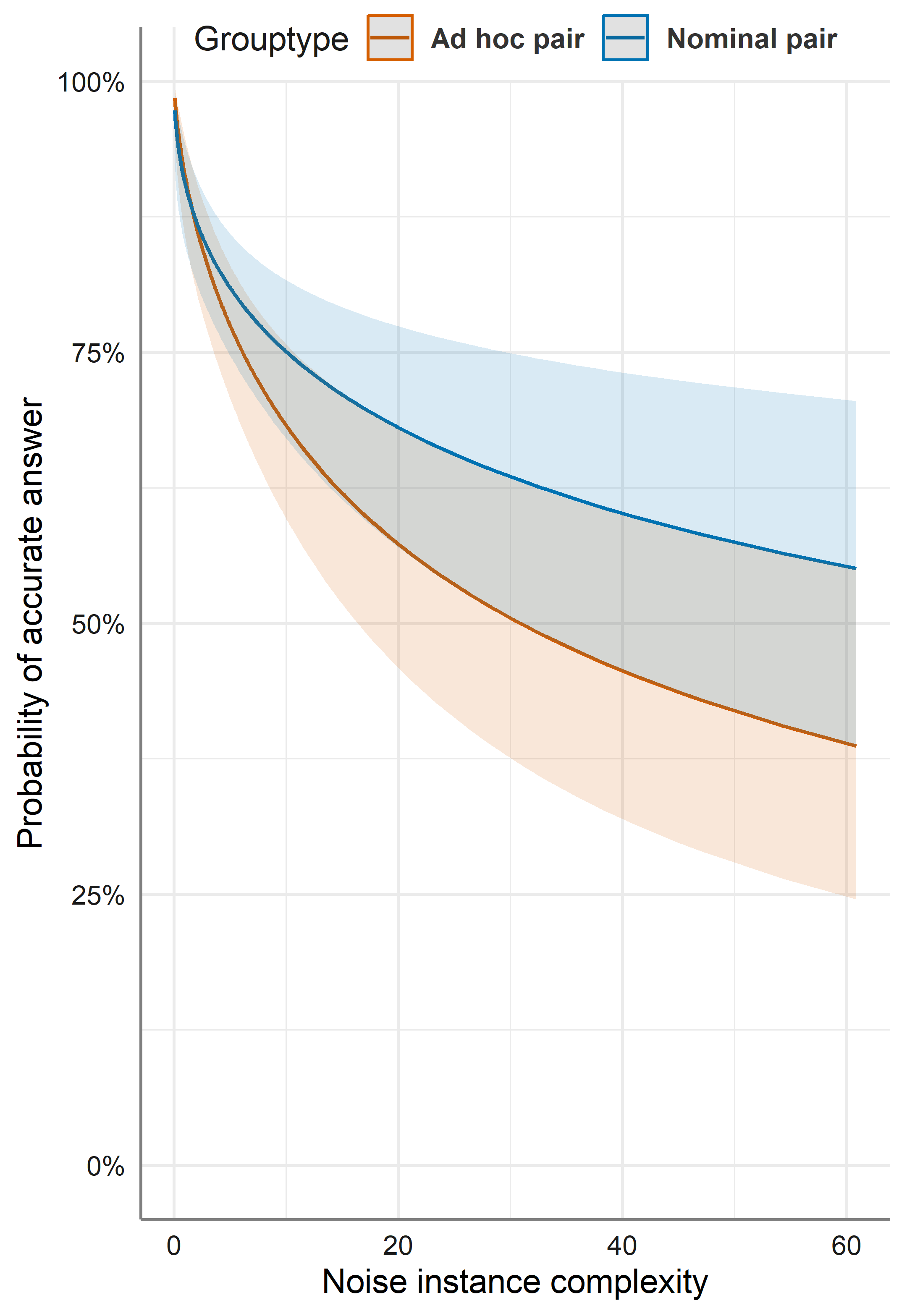} }}%
    \caption{Average predicted probability of an accurate answer for noise instance complexity: ad hoc--individual~(left) and ad hoc--nominal~(right).}%
    \label{fig:marginal_effects_plots_H3}%
\end{figure}

\subsubsection{\textbf{Perceived time duration}}
Immersive analytics~\cite{marriott2018immersive} deals with establishing immersion for analysis' sake, often through a sense of flow, where the very perception of time is distorted. 
For this, participants provided estimates of the time they spent task solving in MR.
In case of several estimates, we used the initial, individual estimate over latter, more educated guesses.
Self-reported and recorded values both followed a log-normal distribution and were transformed.
The two times aligned based on a paired t-test.
Given access to any watches was restricted, this may demonstrate a good blend of realism and engagement.
Pairs and individuals may still perceive the time spent task solving in MR differently, as indicated by a paired t-test~({\small \mbox{$t = 2.282$}, \mbox{$df = 35$}, \mbox{$p = 0.029$}}), which might be caused by an altered in-pair perception~({\small \mbox{$t = -1.929$}, \mbox{$df = 35$}, \mbox{$p = 0.062$}}).

\subsubsection{\textbf{Applied strategies}} 
\begin{figure*}[t]
    \centering
    \subfloat{{\includegraphics[width=2\columnwidth]{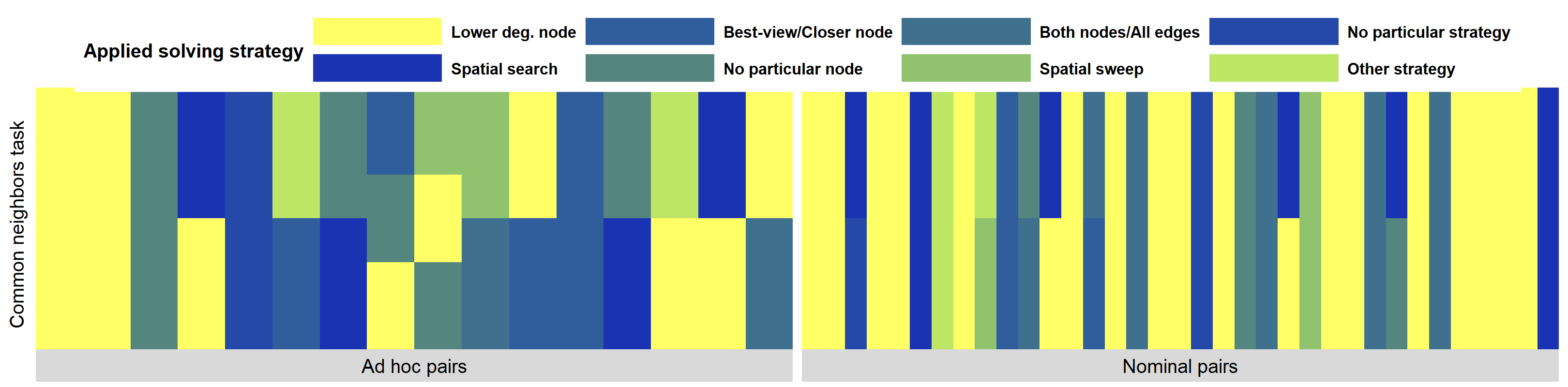} }} \hfill %
    \vspace*{-0.425cm}
    \hspace*{-0.3cm}\subfloat{{\includegraphics[width=2\columnwidth]{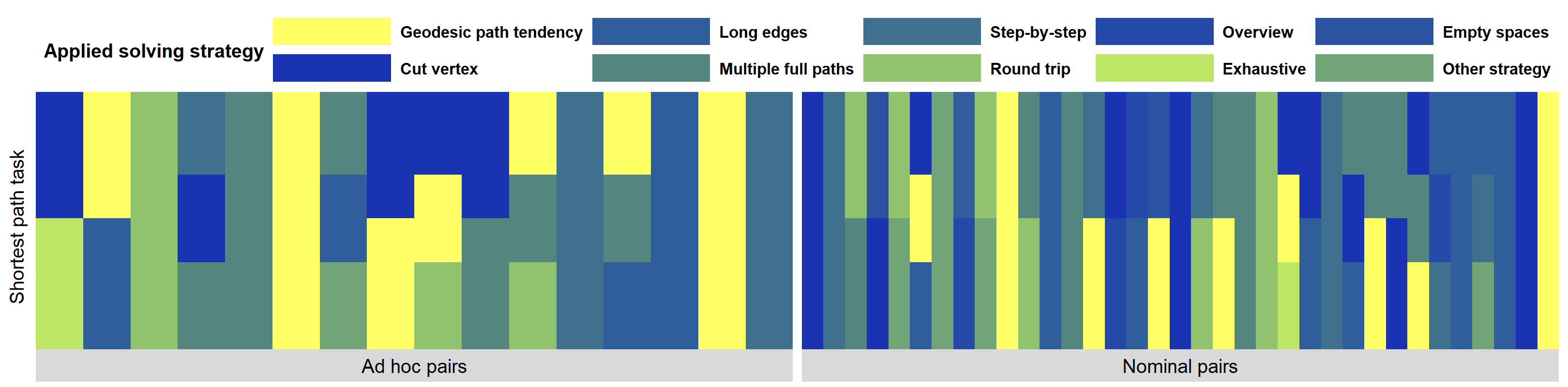} }}%
    \caption{Applied strategy to solve each of the two tasks by the ad hoc and nominal pairs. 
    Vertical stacking in each colormap indicates whether several strategies were used jointly or when these may have evolved gradually~(bottom to top).
    A description of the applied strategies can be found in the Suppl.~Mat.~\cite{darus-4231_2024}.}%
    \label{fig:stratheat}%
\end{figure*}

The strategies applied by the participants are summarized in Fig.~\ref{fig:stratheat}.
In the first task~(CN), a smaller number of participants reported applying an exhaustive search method, such as following all edges or sweeping the space inside the ROI.
Slightly more participants focused on one of the two nodes depending on the viewing perspective, closeness to the participant or for no particular reason.
The majority of participants initiated their search from the lower-degree \mbox{node---the} most efficient strategy topologically~(see~Sec.\;\ref{task1}).
In the second task~(SP), the par\-tic\-i\-pants reported often applying strategies jointly or in succession;~these included focusing on cut vertices or following long edges in the layout, geodesic path following~(see Sec.\;\ref{task2}), and comparing multiple full paths.
Some of the other less commonly applied strategies included step-by-step exploration to circumvent ''dead ends'' and traversing the path in both directions~(round-trip).

Ad hoc pairs regularly consulted each other and, if needed, reached consensus by explaining the intermediate steps.
Ad hoc pairs developed less strategies than nominal pairs~(Fig.\ref{fig:stratheat}), but could better correct for less useful strategies.
Ad hoc pairs subdivided the space, but shared good viewing perspectives.
Indeed, many of the ad hoc pairs' strategies had a tendency to exploit good viewing perspectives.
On the other hand, some pairs struggled to find a common strategy.

\subsubsection{\textbf{Ad hoc pair interaction}} 
Process loss~\cite{steiner1972group} occurred primarily as 1)~view synchronization and 2)~consensus reaching.
At the start, one member would ensure common ground~\cite{clark1991grounding} for the pair.
Later on, members would share new insights, such as label use or the location of a selected node, any doubts they might have, and ultimately the common strategy.
As the pair got accustomed to solving together, com\-mun\-i\-ca\-tion and strategy use became more efficient.
Members would solve in parallel and opt for brief exchanges, some even communicating through the interface, with deeper interactions saved for more complex task instances.
Others would engage in dialogue throughout, e.\,g., while being side by side or proceeding step by step.
Some others would actively split the workload, e.\,g., while counting CN.
We also observed a build-up of trust such that a member may trust their partner without expecting clarification or continuing to solve the task themselves.
Ad hoc pairs rated overall interaction $4.5$ out of five.

\subsubsection{\textbf{Preferred viewing position}}
Participants' reported either combining exocentric and egocentric exploration or relying exclusively on exocentric viewpoints as their preferred viewing position.
Ad hoc pairs preferred exocentric or a mix of both.
Only a few participants adopted an egocentric view.
Lack of awareness and ''fear of contact'' explain why some preferred, at least initially, to stay outside the graph; the labels' orientation was also mentioned. 
Depending on where the selected nodes were, the preferred outer viewpoint varied and could even happen to be above or below the graph. 
If density was high, some participants reported adopting an egocentric, even \textit{node-centric}, view to reduce clutter.
On the other hand, density was why many disliked being inside the graph~("edges would go in my eyes").
Most participants stayed closer to the graph solving the CN task, and leveraged distance to obtain an overview of the graph in the SP task.
Exocentric exploration was crucial in finding good perspectives in any of the tasks.

\subsubsection{\textbf{Challenges and discomforts}}\label{challenges}
Challenges and discomforts introduce an added difficulty to task solving.
Challenges were partitioned based on if the remarks were predominantly made by the ad hoc pairs or not.
Edge crossings and graph density, along with memory demand and visual clarity were most reported by the individuals.
One individual described graph density as ''mentally and physically demanding'', related to the frequent shifts in perspective. 
In this context, some individuals also mentioned long SP. 
Interactive node selection was often the expressed workaround for high memory demand~(also by few pairs).
Visual clarity comprised sharpness and pixelation, as well as difficulties discerning edges and degrees.
Ad hoc pairs focused on challenges hindering their interaction, such as issues related to labels and graph synchronization.
The difference in views, like around any object, was mentioned:~''You had to walk over there and lose where you are.'', which is in line with other comments by pair members adjusting to MR.

Wearing the HMD, at least one discomfort was indicated by over half of all participants.
Responses included primarily head and eye problems, e.\,g., headache and eye strain.
Few also complained from balance problems, fatigue or frustration.

\section{Discussion}
The empirical results show that collaborating around a given \mbox{3D} data representation does not lead to better problem solving such that the ad hoc pair outperforms both the individual and the nominal pair~(\textbf{{\scriptsize RQ}{\small 1}}).
The assembly bonus effects induced by task instance complexity were modest~(\textbf{{\scriptsize RQ}{\small 2}}), providing new evidence for visual graph analysis and complementing existing research on group problem \mbox{solving~\cite{steiner1972group, kerr2004group, hackman2010group, kameda2022information}}.
Duration may, however, be limited to the minute range based on the record of discomforts; more research is needed to investigate prolonged problem solving with HMD.

Even without tool support, such as interactive \mbox{highlighting,} ad hoc pairs were more accurate than individuals~(\textbf{H1}), but failed to outperform nominal pairs in \textbf{H1} and \textbf{H2}.
Against individuals, the accuracy-time trade-off of ad hoc pairs was evident:~pairs used $1.46$ more time to achieve $4.6$\% higher accuracy~(for a distribution with skewness $-2.39$).
Collaborative effects did emerge, most notably in establishing common ground, correcting for suboptimal strategies, and pooling insights to reach consensus. 
Collaboration was mostly fluid and tended to vary between pairs, e.\,g., in terms of communication style or strategy building.
Yet, an ad hoc pair was still only as accurate as the more accurate of two individuals and as slow as the slower of two individuals.
One explanation could be over-coordination~\cite{dourish1992awareness}, which, as one pair found, can be reduced with the use of local copies in a \textit{semi-distributed collaboration}.
But to have pair's collaborative effects outweigh aggregate effects, more than improved coordination might be required.
The challenges, preferences, and strategies participants pointed to during the study therefore confirm the need for dedicated visualization and interaction techniques in the design of CVE.

To investigate the differences for different degrees of perceivable complexity, we introduced the concept of task instance complexity~(\textbf{H3}).
Somewhat surprisingly to us, the differences between group types scaled well with task instance complexity.
This may indicate that high complexity is required for certain synergistic effects to emerge.
Still, we observed two exceptions where synergistic effects did emerge. 
Ad hoc pairs slowed down $10$\% less than nominal pairs as signal complexity increased.
Collaborating hence helped the slower individual reduce completion time increasingly more.
Increases in noise complexity might have the opposite effect, but more research is needed.
The changes in CW may give us more clues~(\textbf{{\scriptsize RQ}{\small 3}}).
In the sample, the odds of high CW for ad hoc pair members increased with accumulated noise complexity~(\textbf{O1}).
The ad hoc pair's strength, its exchange of information, may become its weakness when noise complexity is high.
While the underlying causes were mostly consistent across group types, such as edge crossings and graph density, in ad hoc pairs these may manifest in increasingly difficult coordination and loss of common ground.
These findings could therefore be relevant to group decision-making in MR and have applications for the analysis of networks in MR, such as the power grid of a city~\cite{joos2025enmrgy}.

\subsection*{On task instance complexity}
This study operationalizes task instance complexity for problem solving in 3D and more research can help refine and expand its use.
Hence, we reexamine task instance complexity independent of group type.
The decrease in the probability of an accurate answer for noise instance complexity~(Fig.~\ref{fig:marginal_effects_plots_H3}) might follow a power law, however, more research is needed.
The decrease could be caused by perceptual factors, such as occlusion and overlaps distorting the typical use of perspectives, higher information load, and uncertainty.
Extreme noise complexity thus may render \mbox{analysis} not much different than guesswork.
Less expected to see was that the probability increased with signal instance complexity, even if this was not statistically significant.
The increase could be attributed to the limits of the study design and to the characteristics of the longer shortest paths, such as the likely presence of cut vertices~(see Sec.~\ref{limitations}).   
Completion time, on the other hand, was affected by both signal and noise instance complexity.

\section{Limitations and future work}~\label{limitations}
The type of group is one limitation, given results might differ for practiced groups~\cite{hill1982group}.
Ad hoc pairs are still preferable over mixed groups and practiced pairs are experimentally less feasible.
Conclusions should translate well to more cohesive small groups in the case when ad hoc pairs exceeded.
Different activities, tasks and task types can aid the generalization of results in CVE, while mitigating the influence of potential confounders, including the size of the group, the communication in the group, and the time spent collaborating, as well as the choice of networks and network layouts.
A natural extension of our study is to allow visual specific interactions with the graph, such as visually grouping vertices, filtering part of the network, or entirely redrawing the graph.
It also remains to be studied whether other types of networks and tasks require different task instance complexity measures.
Though consensus-reaching was enforced, we prioritized external validity by allowing participants to determine the level of collaboration themselves~\cite{dourish1992awareness}.
Different levels of collaboration could be investigated in future studies.
Laboratory studies are also prone to social facilitation~\cite{zajonc1965social} in general.

We noticed a slight divergence between the task difficulty participants reported and the accuracy participants achieved.
To shed more light on this aspect, future studies could explore how time allocation affects group accuracy in collocated MR. 
Due to the inclusion of higher path lengths, solution plurality was not limited.
Having a single solution may contribute towards high task instance complexity.
Certain global graph tasks~(e.\,g., Task~2) may therefore require sufficiently large graphs.
There, singular points, such as cut vertices, could be better controlled for, since otherwise accuracy can somewhat counterintuitively increase with signal instance complexity.

MR is inherently visuospatial.
It links visual attention and spatial reasoning and affects both perception and memory.
In our study, we tried not to attenuate one or the other.
Technical limitations, such as the sharing of spatial anchors without prior verification in Meta's store, should also be noted whereby subtle offsets in tracking were noticeable despite strict procedural routines.  
We are also inadvertently limited by current display technology:~HMD, the predominant stereoscopic technology today, are known for obfuscating awareness and facial cues~\cite{billinghurst2002collaborative, billinghurst2018collaborative}.

\section{Conclusion}
In this paper, we reported on a study investigating collaborative problem solving by means of visual graph analysis in MR.
A visuospatial immersive environment, like the one in this study, has potential for overcoming process loss in small groups.
Nominal groups could represent an added benchmark for empirically validating collaborative applications.
Research into signal and noise instance complexity could reveal new mechanisms of dealing with instance complexity in collaborative analytics.
Our contributions could help basic research and find use in the design of CVE.
In particular, the results could motivate the development of interactive methods and visualizations that reduce coordination loss, increase information flow, and improve collaboration.

\section*{Acknowledgments}
\noindent The authors would want to thank all study participants and all pilot study participants from each respective chair.
The authors warmly thank all reviewers for their invaluable feedback.

\bibliography{2.references.bib}

\end{document}